\documentclass{aa}  

\usepackage{graphicx}
\usepackage{txfonts}
\usepackage{hyperref}
\usepackage{gensymb}

\newcommand{\xmm}{{XMM-Newton}\xspace}

\newcommand{\chandra}{{Chandra}\xspace}

\begin{document} 

\title{The eROSITA X-ray telescope on SRG}

\author{
  P. Predehl\inst{1} \and R. Andritschke\inst{1} \and V. Arefiev\inst{9} \and V. Babyshkin\inst{11} \and O. Batanov\inst{9} \and W. Becker\inst{1} \and H. Böhringer\inst{1} \and A. Bogomolov\inst{9} \and T. Boller\inst{1} \and K. Borm\inst{8}\,\inst{12} \and W. Bornemann\inst{1} \and H. Bräuninger\inst{1} \and M. Brüggen\inst{2} \and H. Brunner\inst{1} \and M. Brusa\inst{15} \and E. Bulbul\inst{1} \and M. Buntov\inst{9} \and V. Burwitz\inst{1} \and W. Burkert\inst{1}\thanks{Left Astronomy} \and N. Clerc\inst{14} \and E. Churazov\inst{9}\,\inst{10} \and D. Coutinho\inst{1} \and T. Dauser\inst{4} \and K. Dennerl\inst{1} \and V. Doroshenko\inst{3} \and J. Eder\inst{1} \and V. Emberger\inst{1} \and T. Eraerds\inst{1} \and A. Finoguenov\inst{1} \and M. Freyberg\inst{1} \and P. Friedrich\inst{1} \and S. Friedrich\inst{1} \and M. Fürmetz\inst{1}\protect\footnotemark[1] \and A. Georgakakis\inst{13} \and M. Gilfanov\inst{9}\, \inst{10} \and S. Granato\inst{1}\protect\footnotemark[1] \and C. Grossberger\inst{1}\protect\footnotemark[1] \and A. Gueguen\inst{1} \and P. Gureev\inst{11}\and F. Haberl\inst{1} \and O. Hälker\inst{1} \and G. Hartner\inst{1} \and G. Hasinger\inst{5} \and H. Huber\inst{1} \and L. Ji\inst{3}\and A. v. Kienlin\inst{1} \and W. Kink\inst{1} \and F. Korotkov\inst{9} \and I. Kreykenbohm\inst{4} \and G. Lamer\inst{7} \and I. Lomakin\inst{11} \and I. Lapshov\inst{9} \and T. Liu\inst{1} \and C. Maitra\inst{1} \and N. Meidinger\inst{1} \and B. Menz\inst{1}\protect\footnotemark[1] \and A. Merloni\inst{1} \and T. Mernik\inst{12} \and B. Mican\inst{1} \and J. Mohr\inst{6} \and S. Müller\inst{1} \and K. Nandra\inst{1} \and V. Nazarov\inst{9} \and F. Pacaud\inst{8} \and M. Pavlinsky\inst{9} \and E. Perinati\inst{3} \and E. Pfeffermann\inst{1} \and D. Pietschner\inst{1} \and M. E. Ramos-Ceja\inst{1} \and A. Rau\inst{1} \and J. Reiffers\inst{1} \and T.H. Reiprich\inst{8} \and J. Robrade\inst{2} \and M. Salvato\inst{1} \and J. Sanders\inst{1} \and A. Santangelo\inst{3} \and M. Sasaki\inst{4} \and H. Scheuerle\inst{12}\protect\footnotemark[1] \and C. Schmid\inst{4}\protect\footnotemark[1] \and J. Schmitt\inst{2} \and A. Schwope\inst{7}  \and A. Shirshakov\inst{11} \and M. Steinmetz\inst{7} \and I. Stewart\inst{1} \and L. Strüder\inst{1}\protect\footnotemark[1] \and  R. Sunyaev\inst{9}\, \inst{10}  \and C. Tenzer\inst{3} \and L. Tiedemann\inst{1}\protect\footnotemark[1] \and J. Trümper\inst{1} \and V. Voron \inst{16} \and P. Weber \inst{4} \and J. Wilms \inst{4}  \and V. Yaroshenko\inst{1} 
}

\institute{
     Max-Planck-Institut f\"ur extraterrestrische Physik,
     Gie{\ss}enbachstra{\ss}e, D-85748 Garching, Germany
\and Universit\"at Hamburg, Hamburger Sternwarte, 
               Gojenbergsweg 112, D-21029 Hamburg, Germany
\and Institut f\"ur Astronomie und Astrophysik, Abteilung Astronomie,
               Universit\"at T\"ubingen, Sand 1, D-72076 T\"ubingen, Germany
\and Universit\"at Erlangen/Nürnberg, Dr.-Remeis-Sternwarte, 
     Sternwartstra{\ss}e 7, D-96049, Bamberg, Germany
\and ESAC Camino bajo de Casillo S/N
     Villanueva de la Canada (Madrid), Spain
\and Ludwig-Maximilians-Universit\"at München, Universit\"atssternwarte,
                Scheinerstra{\ss}e 1, D-81679 Munich, Germany
\and Leibniz Institut f\"ur Astrophysik Potsdam, 
     An der Sternwarte 16, D-14482 Potsdam, Germany
\and Argelander-Institut f\"ur Astronomie, Universit\"at Bonn,  
     Auf dem H\"ugel 71, D-53121 Bonn, Germany
\and IKI, Space Research Institute, 84/32 Profsouznaya ulitsa, 
     Moscow 117997, Russian Federation
\and Max-Planck-Institut f\"ur Astrophysik, Karl-Schwarzschild-Stra{\ss}e, D-85741 Garching, Germany 
\and Lavochkin Association, 24 Leningradskaya ulitsa, 
     Khimki 141400, Moscow Region, Russian Federation
\and Deutsches Zentrum für Luft- und Raumfahrt, 
     K\"onigswinterer Stra{\ss}e. 522-524, D-53227 Bonn, Germany
\and National Observatory of Athens,V. Paulou \& I. Metaxa, Athens, Greece
\and Institut de Recherche en Astrophysique et Planétologie (IRAP), Universit{\'e} de
     Toulouse, CNRS, UPS, CNES, F-31028 Toulouse, France
\and Dipartimento di Fisica e Astronomia, Alma Mater Studiuorum Universit\`a di Bologna,      via Gobetti 93/2,  40129 Bologna, Italy and INAF - Osservatorio di Astrofisica e        Scienza dello Spazio  di Bologna, via Gobetti 93/3,  40129 Bologna, Italy
\and State Space Corporation Roscosmos, 42 Schepkina ulitsa 42, Moscow 107996, Russian Federation
}

\date{Received May 15, 1899; accepted May 16, 2029}

 

  \abstract{eROSITA (extended ROentgen Survey with an Imaging Telescope Array) is the primary instrument on the  Spectrum-Roentgen-Gamma (SRG) mission, which was successfully launched on July 13, 2019, from the Baikonour cosmodrome. After the commissioning of the instrument and a subsequent calibration and performance verification phase, eROSITA started a survey of the entire sky on December 13, 2019. 
  By the end of 2023, eight complete scans of the celestial sphere will have been performed, each lasting six months.  At the end of this program, the eROSITA all-sky survey in the soft X-ray band (0.2--2.3\,keV) will be about 25 times more sensitive than the ROSAT All-Sky Survey, while in the hard band (2.3--8\,keV) it will provide the first ever true imaging survey of the sky. 
  The eROSITA design driving science is the detection of large samples of galaxy clusters up to redshifts $z>1$ in order to study the large-scale structure of the universe and test cosmological models including Dark Energy.
  In addition, eROSITA is expected to yield a sample of a few million AGNs, including obscured objects, revolutionizing our view of the evolution of supermassive black holes. The survey will also provide new insights into a wide range of astrophysical phenomena, including X-ray binaries, active stars, and diffuse emission within the Galaxy.
  Results from early observations, some of which are presented here, confirm that the performance of the instrument is able to fulfil its scientific promise.
  With this paper, we aim to give a concise description of the instrument, its performance as measured on ground, its operation in space, and also the first results from in-orbit measurements.}


\maketitle

\section{Introduction}

The eROSITA (extended ROentgen  
Survey with an Imaging Telescope Array) instrument concept is based on a long series of previous scientific and technological developments at the Max Planck Institute for extraterrestrial Physics (MPE), 
dating back to the very successful German/US/UK ROSAT X-ray satellite mission
\citep[1990-1999;][]{Truemper1982}, which was developed and managed under the leadership of MPE. ROSAT carried out the
first complete survey of the sky with an imaging X-ray telescope in the energy range between 0.1 and 2.4 keV,
and performed tens of thousands of pointed observations. Just as ROSAT has been the reference for the past 30
years, so will eROSITA on SRG (Spectrum-Roentgen-Gamma) be the reference in the future.

The SRG is an
astrophysical observatory, comprising two imaging X-ray telescopes: the primary payload eROSITA, developed under the responsibility of MPE, Germany, and 
ART-XC (Astronomical Roentgen Telescope X-ray Concentrator), an X-ray mirror telescope complementing the eROSITA sensitivity towards higher energies, developed under the lead of the Russian Space Research Institute IKI \citep{Pavlinsky2018}. 
The scientific payloads of SRG are mounted on the ``Navigator'' spacecraft platform built by Lavochkin Association (``NPOL'') in Khimky near Moscow in Russia. Navigator has been developed as a universal medium-class platform for scientific missions to be launched into various orbits. Since January 2011, the Navigator platform has been used in the three \mbox{Elekro-L} meteorological satellite missions, as well as in the scientific \mbox{Spektr-R} mission \citep[\mbox{RadioAstron};] []{RadioAstron2013}, which was launched in 2011 and operated until 2018.

This paper presents a concise description of the main scientific goals of eROSITA, of the instrument itself, of its performance as measured on ground, and its operations in space, and presents some of the first results from in-orbit measurements. More details about the in-flight calibration program, and the performance of the various eROSITA subsystems, as well as a description of the ART-XC telescope and of the SRG mission as a whole, will be published separately.
\section{The eROSITA mission}
\subsection{Scientific objectives}

eROSITA was designed as a sensitive wide-field X-ray telescope capable of delivering deep, sharp images over very large areas of the sky. The advantages of wide-field X-ray telescopes have been discussed for decades \citep[see e.g.,][]{Burrows1992}. However,
most of the current generation of sensitive focusing X-ray telescopes, including the flagship observatories \chandra (NASA) and \xmm (ESA), have a relatively small field of view, making it difficult to map large volumes of the Universe in a short amount of time. For this reason, wide-area (or all-sky) surveys in X-rays tend to be limited to the very brightest subset of the population, that is, mostly nearby sources in the Milky Way. The notable exception is the ROSAT all-sky survey, which was performed over six months in 1990 and at the time increased the number of known X-ray sources by a factor 20 \citep{Truemper1993,Voges1999,Boller2016, Salvato2018}.

A deep view of the X-ray sky over large areas gives unique insights into the cosmological evolution of large-scale structure. On the one hand, the diffuse plasma that virializes within the most massive dark matter halos heats up to temperatures of tens of millions of degrees, leading to copious emission of X-ray photons \citep{Bahcall1977,Cavaliere1978,Sarazin1986,Rosati2002,vo05,a05,n05,b06,bk09,aem11,rbe13}.
On the other hand, X-ray emission is a universal signature of accretion of matter onto the supermassive black holes (SMBHs) that likely seed the entire population of galaxies and may strongly influence their formation and subsequent evolution \citep{Hopkins2008,Hickox2009,Fabian2012,Alexander2012,Kormendy2013,Brandt2015}. Thus, a sufficiently detailed map of the Universe in X-rays highlights both the interconnected structure of the dark-matter web and the role of black holes in galaxy formation. The required sensitivity of an all-sky survey that could map the large-scale structure implies moreover that data are accumulated for a large variety of astronomical source classes, and for a plethora of possible science applications well beyond the main design-driving objectives. These data are therefore endowed with tremendous legacy value.

In the soft X-ray band (0.2--2.3\,keV), the eROSITA survey was designed to be about 25 times more sensitive than the ROSAT all-sky survey, while in the hard band (2.3--8\,keV) it will provide the first ever true imaging survey of the sky at those energies. With soft X-ray effective area and on-axis spatial resolution comparable to \xmm, better energy resolution, and a much larger field of view, eROSITA is a powerful X-ray telescope. Together with ART-XC, which expands the energy range up to 30 keV, this makes SRG a highly competitive X-ray observatory.

According to several independent analyses \citep{Pillepich2012, Merloni2012, Kolodzig2013,Borm2014, Pillepich2018, clerc2018,Zandanel2018,Comparat2019},
eROSITA is expected to yield a sample of at least 100\,000 clusters of galaxies, a few million active galactic nuclei (AGNs), and around 700\,000 active stars among many other X-ray-emitting objects within and outside our own Galaxy. 

Moreover, such a deep imaging survey at medium to high spectral resolution, with its scanning strategy that is sensitive to a range of variability timescales from tens of seconds to years (see Sect.~\ref{sec:planning} below), will undoubtedly open up a vast discovery space for the study of rare, unexpected, or even yet unpredictable high-energy astrophysical phenomena \citep{Merloni2012,Khabibullin2014}.

The scientific exploitation of the eROSITA all-sky survey data is shared equally between a German and a Russian consortium. Two hemispheres
\footnote{Data rights are split by galactic longitude ($l$); data with $l<180 \degree$ belong to the Russian consortium, 
while data with $l>180 \degree$ belong to the German consortium.}
of the sky have been defined, over which each team has unique scientific data exploitation rights, while all-sky data are available to both teams for purposes of  technical and calibration analysis, pipeline validation, and so on.  This simple scheme guarantees a fair share of both Galactic and extragalactic areas. A collaboration between the two consortia is encouraged particularly for those kinds of science which require the full sky for their exploitation.

Results from observations taken during the Calibration and Performance Verification (Cal-PV) phase of the mission, as well as from the first all-sky survey, in part briefly outlined in this work, confirm that these expectations will be fulfilled.

\begin{table*}
\caption{Major mission milestone for eROSITA since the launch of SRG.}   \label{tab:milestones}      
\begin{tabular}{ll}     
\hline                    
Date & Event \\
\hline                    
2019/7/13, 15:31 MSK$^+$ & Launch. \\
2019/7/13, 17:31 MSK$^+$ & Insertion into L2 trajectory and separation from Block-DM03 space tug. \\
2019/7/13, 17:39 MSK$^+$ & First eROSITA telemetry received. \\
2019/7/22 & First trajectory correction maneuver. \\
2019/7/23 & Telescope cover opens. Outgassing period begins. \\
2019/8/06 & Second trajectory correction maneuver. \\
2019/8/22 & Camera cool-down. Start of camera commissioning. \\
2019/9/15 & Commissioning first light with 2 TMs$^*$ [LMC]. \\
2019/10/16-18 & First light with all 7 TMs$^*$ [LMC]. End of extended commissioning. \\
2019/10/18 & Calibration and performance verification program begins. \\
2019/10/21 & Third trajectory correction maneuver and insertion into L2 Halo orbit. \\
2019/12/8 & Calibration and performance verification program ends. \\
2019/12/13 & All-sky survey begins.\\
\hline 
$^+$MSK: Moscow time (= UTC + 3hr). \\
$^*$TMs: Telescope modules.
\end{tabular}
\end{table*}
\subsection{Major mission milestones}

Table~\ref{tab:milestones} presents the sequence of the major mission milestones for eROSITA from its launch to the completion of the Cal-PV program and the start of the first all-sky survey. The SRG was launched on July 13, 2019, at 15:31 Moscow time from Baikonur, Kazakhstan, using a Proton-M rocket and a BLOK DM-03 upper stage. On its three months cruise to the second Lagrangian point (L2) of the Earth--Sun system, 1.5 million km in the anti-sun direction, the spacecraft and instruments underwent commissioning, checkout, calibration and an early science performance verification program. Since mid-October 2019, SRG is in a six-month-periodic halo orbit around L2, with a major semiaxis of about 750\,000~km within the ecliptic plane and about 400\,000~km perpendicular to it. 

Since December 13, 2019, the mission has been in its survey mode, with the spacecraft engaged in a continuous rotation around an axis pointing to the neighborhood of the Sun. Since the whole sky is covered every half year (see section~\ref{sec:planning}), a total of eight scans will be completed after the planned four years of survey-mode operations. The all-sky survey program will be followed by a phase of pointed observations, expected to start in late 2023 or early 2024, including access through regular announcements of opportunity for the worldwide astrophysical community.

\section{The eROSITA instrument}

\begin{figure}[h]
\includegraphics{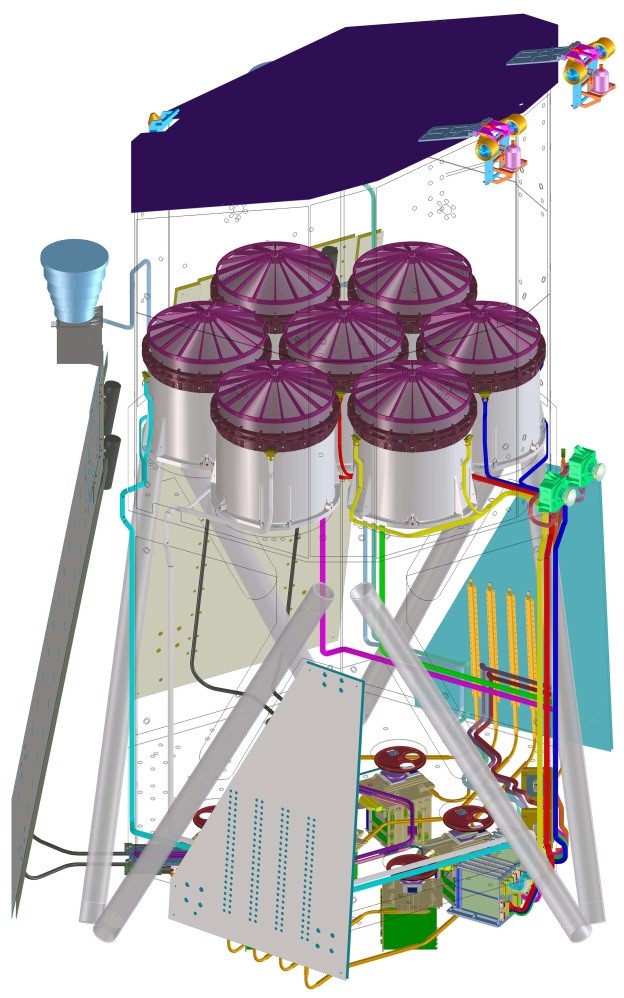}
\caption{Schematic view of the eROSITA telescope with (from top): Front cover, 7 MAs (baffle, mirror, electron deflector), and 7 CAs (filter wheel, camera, electronics box). In addition there are two star trackers (one seen) and four large radiators to cool the cameras and their electronics. The telescope structure is lightly shaded.}
\label{scheme}
\end{figure}

\begin{figure}
\centering\includegraphics [width = 9cm]{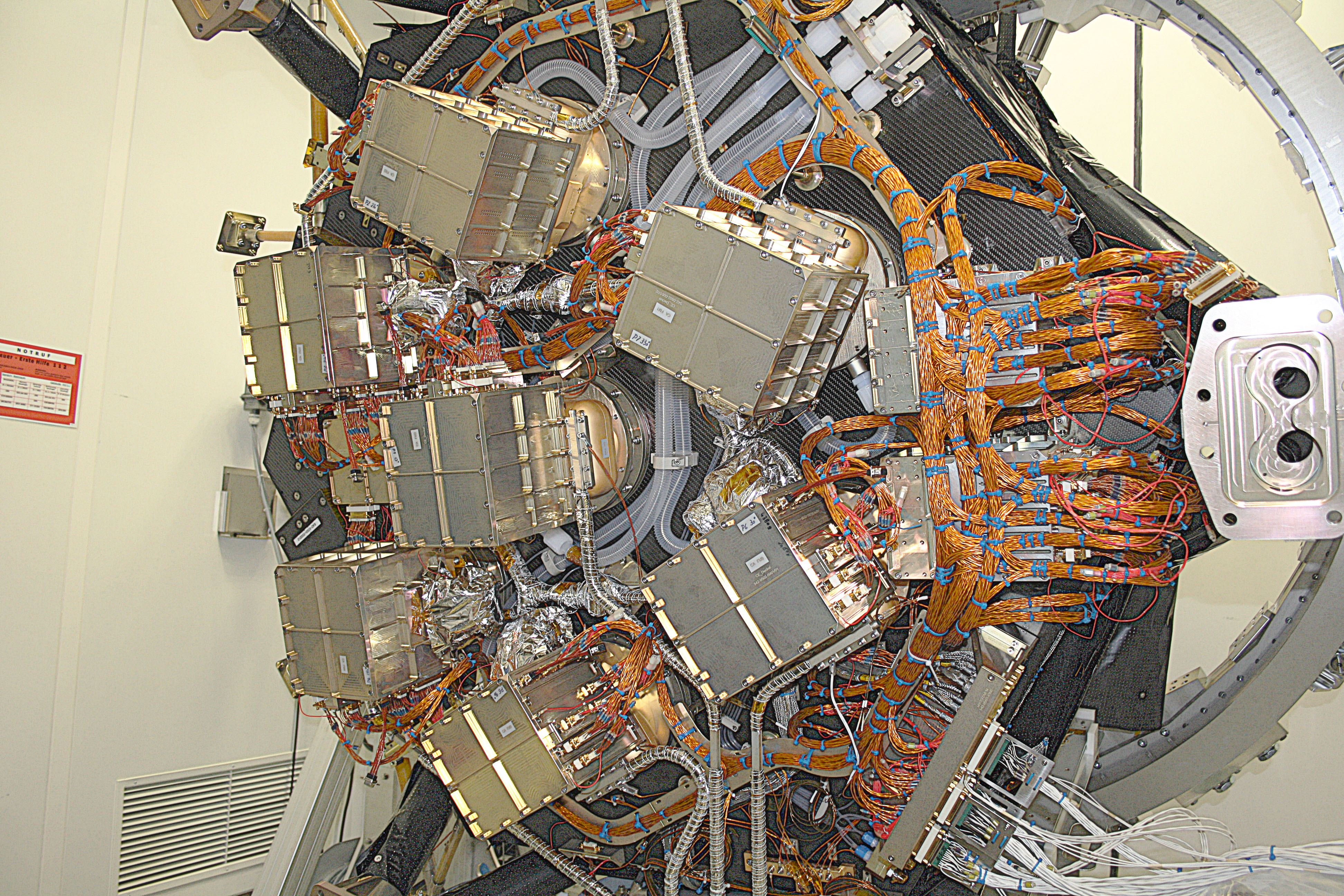}
\caption{Rear view of the telescope with all seven CAs  installed. Each CA consists of a filter wheel, a camera,  and a camera electronics box. Also seen are heat pipes, purging tubes, and a harness.}
\label{cameras}
\end{figure}

Figure \ref{scheme} shows a schematic view of the telescope. eROSITA consists of seven identical and co-aligned X-ray mirror assemblies (MAs) housed in a common optical bench. The supporting telescope structure consists of a system of carbon-fibre honeycomb panels connecting the seven MAs on one side with the associated seven camera assemblies (CAs) on the other side. A hexapod structure forms the mechanical interface to the S/C bus. The seven individual telescope modules (TMs) are arranged in a hexagonal shape
\citep{Eder2018}; see also Figures \ref{mirrors} and \ref{cameras}. 

Two star sensors (Sodern SED26) are mounted on eROSITA. They belong to the attitude system of the S/C but serve also for  determination of the boresight. They are read out once per second, and the specified accuracy is 3 arcsec (3$\sigma$).

The dimensions of the telescope structure are approximately 1.9\,m (diameter) $\times$ 3.2\,m (height in launch
configuration, with closed front cover). The total weight of eROSITA is 808\,kg. Table~\ref{BIP-1} shows the 
basic eROSITA instrument parameters in launch configuration.

\begin{table*}[h]
\caption{Basic instrument parameters in launch configuration}             
\label{BIP-1}      
\centering          
\begin{tabular}{|c c|c c|c c|}     
\hline                    
\multicolumn{2}{|c|}{Instrument} & \multicolumn{2}{|c|}{7 Mirror assemblies} & \multicolumn{2}{|c|}{7 Camera assemblies}\\
\hline                    
Size & 1.9\,m $\diameter$ $\times$ 3.5\,m & Diam. of outer shell   & 358\,mm    & CCD image  & $2.88 \times 2.88\,\mathrm{cm}^2$  \\  
     Mass & 808\,kg              & Number of shells       & 54         & Pixel size   & $75\mu\mathrm{m}\times 75\mu\mathrm{m}$ \\
     Power & 522\,W max.         & Focal length           & 1600\,mm   & Time Resol. & 50 ms \\
     Data rate & 600\,MB/day max. &                        &            &             &      \\
\hline                  
\end{tabular}
\end{table*}

\subsection{eROSITA mirror assemblies}

\begin{figure}
\centering\includegraphics [width = 9cm]{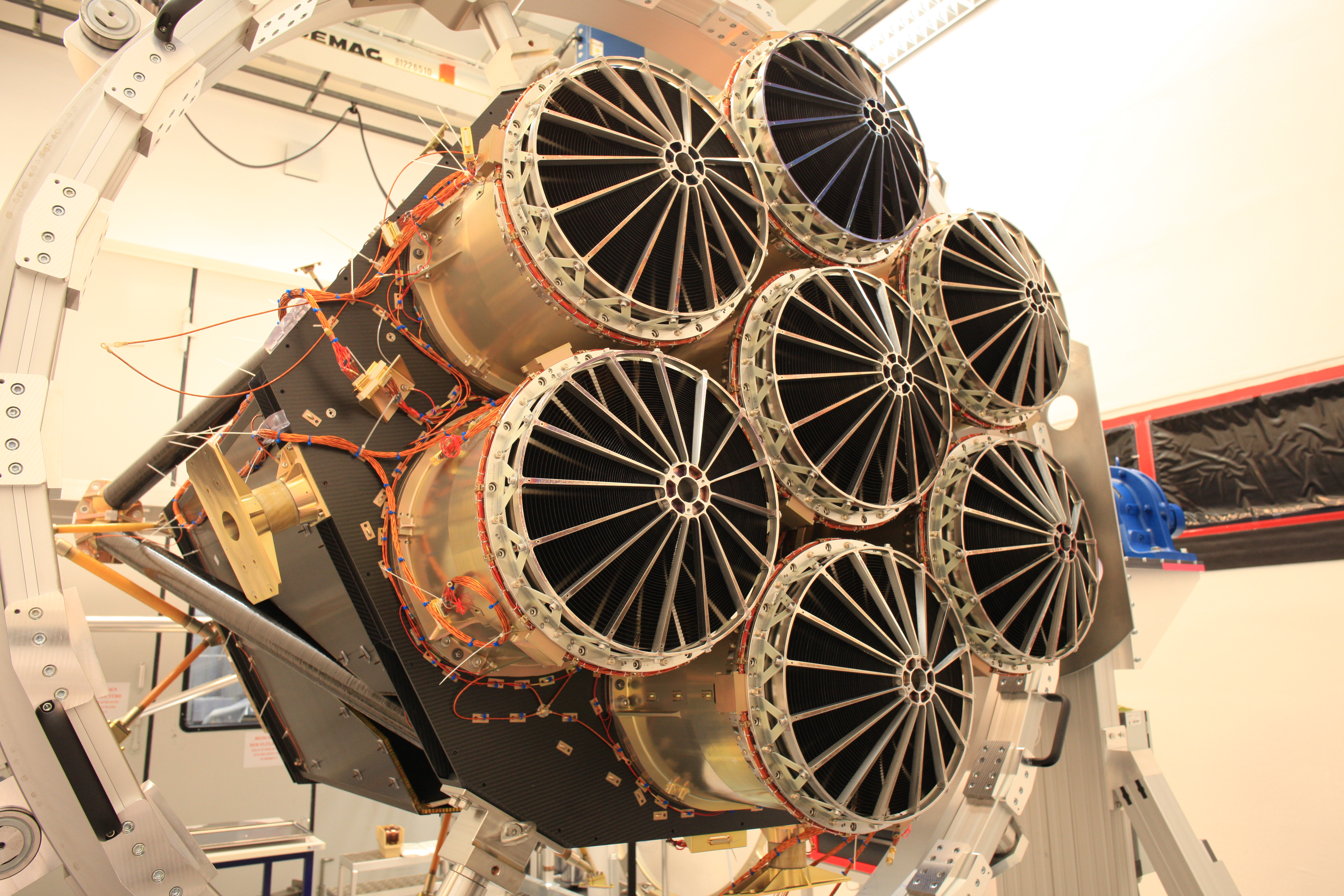}
\caption{Front view of the telescope with all seven MAs installed. Each MA consists of a mirror module, an X-ray baffle in front, and a magnetic electron deflector behind.}
\label{mirrors}
\end{figure}

Each of the 
mirrors comprises 54 paraboloid/hyperboloid mirror shells in a Wolter-I geometry, with an outer diameter of 360 mm and a common focal length of 1\,600 mm (Fig. \ref{mirrors}, \citealt{Friedrich2008}; \citealt{Arcangeli2017}). The mirror shells consist of electroformed nickel with gold evaporated on the reflecting side.
The average on-axis resolution of the seven 
MAs as measured on ground is $16.1''$ half-energy width (HEW) at 1.5\,keV (see Table \ref{psfs} for individual values). 
The unavoidable off-axis blurring typical of Wolter-I optics is compensated by a 0.4~mm shift of the cameras towards the mirrors. This puts each telescope out of focus, leading to a slight degradation of the on-axis performance (about $18''$), but improved angular resolution averaged over the field of view (about $26''$). 

The principle  of a Wolter-I mirror system cannot prevent photons from X-ray sources outside the field of view reaching the camera by single reflection on the hyperboloid. This X-ray stray light has in general the effect of increasing the background, but the distortion of the X-ray image can be even more dramatic if there are bright sources just outside the field of view. The unwanted photons can be suppressed using an X-ray baffle placed in front of the mirror module. Due to the short focal length,  a system of sieve plates, as on XMM-Newton, unfortunately does not work. Instead, the eROSITA X-ray baffle consists of 54 concentric invar cylinders mounted on spider wheels, precisely matching the footprint of the parabola entrance of each mirror shell \citep{Friedrich2014}. Magnetic electron deflectors behind the mirrors help to further reduce the background due to low-energy cosmic-ray electrons and complete the MAs.

\subsection{eROSITA camera assemblies}

\begin{figure*}[t]
\centering\includegraphics [width = 18cm]{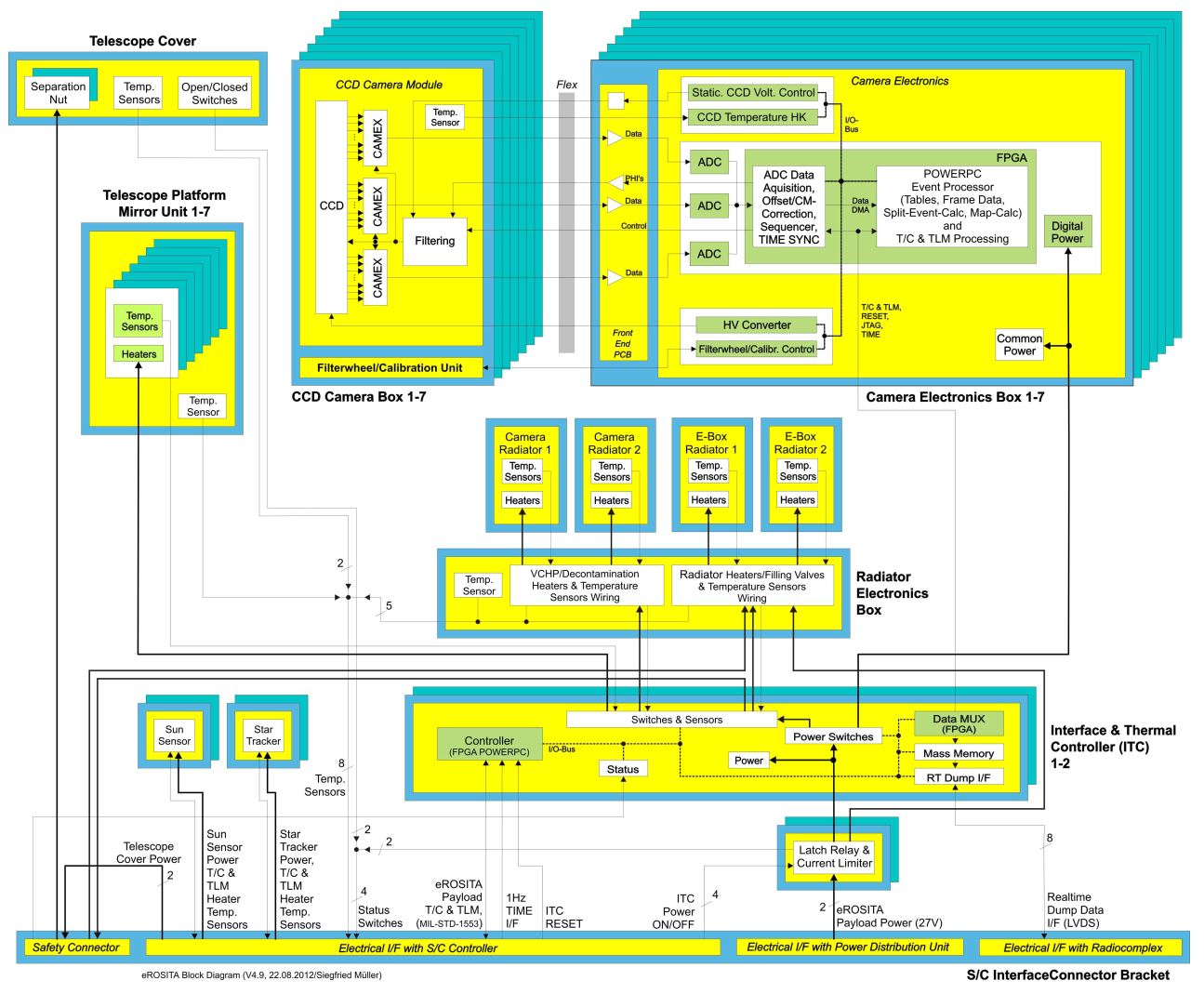}
\caption{Functional schematic of eROSITA electronics.}
\label{electronics}
\end{figure*}

Each MA has a charge-coupled-device (CCD) camera in its focus \citep{Meidinger2014}. The eROSITA CCDs each have $384\times384$ pixels in an image area of $28.8\,\mathrm{mm} \times 28.8\,\mathrm{mm}$.
The circular field of view with a diameter of $1\fdg{}03$ limited by an aperture stop is
exactly contained within this square active CCD area. 
Each pixel corresponds to a sky area of $9\farcs{}6 \times 9\farcs{}6$. The nominal integration time of the eROSITA CCDs is 50\,ms. 

Unlike the pnCCD camera on \xmm \citep{Strueder2001}, the eROSITA CCDs contain a framestore area. The image area is shifted within 0.115\,ms into this framestore area and the read-out happens within 9.18\,ms. This scheme substantially reduces  the amount of so-called ``out-of-time'' events, which are recorded during read-out. In the case of XMM EPIC-pn, this results in prominent artifacts in the images known as the read-out streaks, which are particular evident when bright point sources are observed. This effect is almost entirely suppressed via this frame-store design in the eROSITA CCDs.

The CCDs are protected against cosmic particle radiation by means of a massive copper shield. X-ray fluorescence radiation generated by cosmic particles is reduced by a graded shield consisting of aluminium, beryllium, and/or boron-carbide.

For calibration purposes, each camera has its own filter wheel with four positions: (1) OPEN, primarily used for outgassing. (2) FILTER, which is the standard observing mode. The filter consists of a polyimide foil with an aluminium layer as light filter for cameras 5 and 7, while cameras 1, 2, 3, 4, and 6 have the aluminium directly deposited on the CCD. For these cameras the polyimide foil acts as a contamination shield.
 (3) CLOSED, for measuring instrumental background.
 (4)) CALIB, with a radioactive $^{55}$Fe source and an aluminium/titanium target providing three spectral lines at 5.9\,keV (Mn-K$\alpha$), 4.5\,keV (Ti-K$\alpha$) and 1.5\,keV (Al-K). 

For optimal performance during operations, the CCDs are cooled down to about $-85^\circ$C by means of passive elements.
To cool the cameras, a complex system of cryogenic heat pipes had to be developed. This system comprises seven small camera heat pipes which are connected to two redundant ring heat pipes, which in turn are connected to two large radiators by means of four so-called sVCHPs (``switchable variable conductance heat pipes''). A VCHP provides some temperature stability by itself. We added a mechanism (``switch'') in order to prevent the heat pipes from working so that the cameras remain warm for outgassing during the first days
after launch \citep{Fuermetz2008}. 

\subsection{eROSITA electronics}

\begin{figure}[h]
\centering\includegraphics [width=9cm]{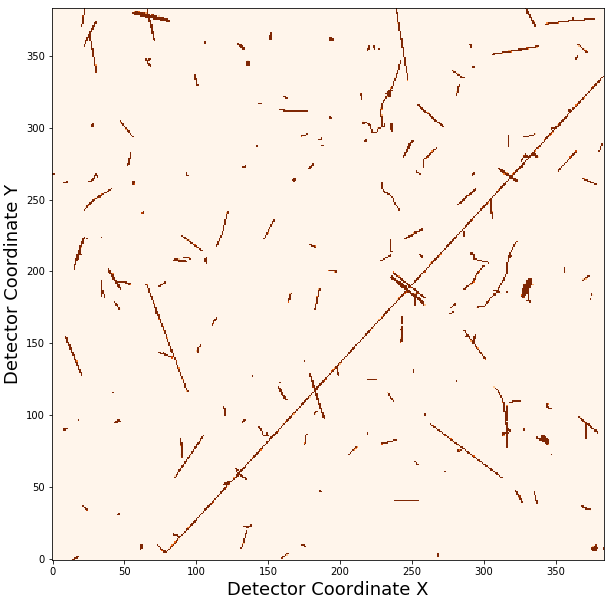}
\caption{Stack of numerous CCD raw frames from all TMs taken in orbit. This shows the variety of cosmic radiation particles hitting the cameras at all times, which  in orbit amount to typically one to three per readout frame every 50 ms.}
\label{fig:mips}
\end{figure}

The electronics for onboard processing of the camera data consists of seven sets of camera electronics (CE), each one mounted and interfacing to the cameras through a flexlead (Fig. \ref{electronics}). At the heart of each CE is a Xilinx Virtex IV \textit{Field Programmable Gate Array} with an embedded PowerPC processor. Each of the CEs provides the proper voltage control and readout timing of the associated camera, and performs the onboard data processing within the time constraints of the camera integration time. Following the correction of offsets and common mode disturbances, the event signals are extracted.
The processing of events first comprises the rejection of signals caused by cosmic particles (Fig. \ref{fig:mips}): while rare on ground, there are on average one to three particles per readout frame in orbit. Finally, together with housekeeping data, events are  coded into telemetry frames.

Interfacing with the CEs is the interface and thermal controller (ITC). This unit receives the telemetry generated by each CE and stores it in the mass memory, commands each of the CEs, and controls the power distribution. The ITC also houses the thermal control hardware and software, regulating the temperatures of the seven mirror modules to $20\pm 0.25^\circ$C and CCD detectors to 
$-84.75\pm 0.25^\circ$C. In addition, the interface to the spacecraft is provided through the ITC, where either the real-time camera data or the full telemetry stored in the mass memory are streamed through to the ground. A MIL1553 interface to the spacecraft is used for commanding eROSITA through the ITC and to downlink a subset of the instrument’s housekeeping. Given its criticality, the ITC is a cold redundant unit \citep{Coutinho2018}.

\begin{table*}
\caption{Energy resolution [eV] and QEs of the eROSITA CAs as measured on ground.
``QE12346'' is the QE for cameras TM1, 2, 3, 4, and 6, which have filters directly deposited onto the CCD; ``QE57'' is the QE of cameras TM5 and TM7, which have their filters in the filter wheel. }             \label{ergres}      
\centering          
\begin{tabular}{l c c c c c c c c c c}     
                                   & TM1 & TM2 & TM3 & TM4 & TM5 & TM6 & TM7 & QE12346 & QE57 &  \\
\hline                                                                 
    C-K  at 0.277\,keV           &  58$\pm$0.3 &  58$\pm$0.3 &  58$\pm$0.4 &  58$\pm$0.3 &  50$\pm$0.2 &  58$\pm$0.4 &  49$\pm$0.2 & 12.4$\pm$1.7\,\%  &  31.3$\pm$4.4\,\% \\  
    O-K  at 0.525\,keV           &  64$\pm$0.2 &  65$\pm$0.3 &  66$\pm$0.3 &  64$\pm$0.2 &  57$\pm$0.3 &  63$\pm$0.2 &  56$\pm$0.4 & 42.2$\pm$1.6\,\%  &  51.3$\pm$2.1\,\% \\  
    Cu-L at 0.93\,keV            &  70$\pm$0.3 &  74$\pm$0.3 &  72$\pm$0.3 &  70$\pm$0.3 &  68$\pm$0.3 &  70$\pm$0.3 &  68$\pm$0.3 & 80.0$\pm$4.5\,\%  &  83.2$\pm$4.7\,\% \\  
    Al-K at 1.49\,keV            &  77$\pm$0.3 &  82$\pm$0.3 &  80$\pm$0.3 &  77$\pm$0.3 &  75$\pm$0.3 &  77$\pm$0.3 &  77$\pm$0.2 & 94.0$\pm$4.1\,\%  &  94.8$\pm$4.2\,\% \\  
    Ti-K$\alpha$ at 4.51\,keV    & 118$\pm$0.5 & 125$\pm$0.6 & 122$\pm$0.6 & 118$\pm$0.6 & 116$\pm$0.6 & 118$\pm$0.6 & 117$\pm$0.6 & 97.9$\pm$2.2\,\%  &  98.2$\pm$2.2\,\% \\
    Fe-K$\alpha$ at 6.40\,keV    & 138$\pm$0.6 & 145$\pm$0.7 & 142$\pm$0.7 & 138$\pm$0.6 & 135$\pm$0.7 & 138$\pm$0.7 & 136$\pm$0.7 & 98.9$\pm$2\,\%  &  99.1$\pm$2\,\% \\
    Cu-K$\alpha$ at 8.04\,keV    & 158$\pm$0.7 & 167$\pm$0.7 & 163$\pm$0.7 & 159$\pm$0.7 & 155$\pm$0.6 & 159$\pm$0.6 & 156$\pm$0.7 & 99.3$\pm$2\,\%  &  99.4$\pm$2\,\% \\
    Ge-K$\alpha$ at 9.89\,keV    & 178$\pm$1.0 & 181$\pm$1.0 & 182$\pm$1.1 & 173$\pm$1.1 & 170$\pm$1.0 & 174$\pm$1.1 & 175$\pm$1.0 & 96.9$\pm$2\,\%  &  96.9$\pm$2\,\% \\
\hline                           
\end{tabular}
\end{table*}

\section{Ground calibration}

\subsection {Camera calibration}

The on-ground calibration measurements of the eROSITA CAs started in December 2015 and lasted until June 2016. They were all performed at the PUMA X-ray test facility at MPE. More than 400 million events were recorded for each camera, producing data sets of excellent statistical quality.

\begin{figure}
\includegraphics[width=9cm]{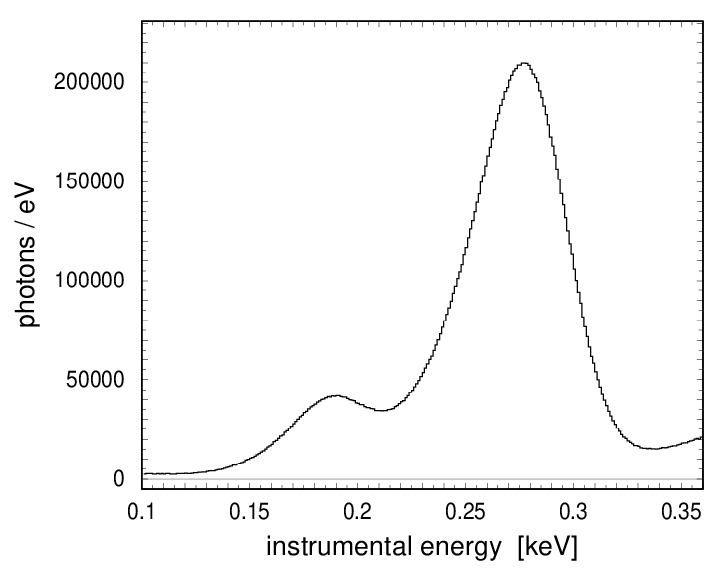}
\caption{Spectrum of boron and carbon emission lines taken with camera of TM7. This demonstrates the excellent low-energy response and resolution of the eROSITA cameras, about 49\,eV FWHM at these energies.
}
\label{Boron}
\end{figure}

The energy calibration was derived for a global 
threshold of 46 adu ($\sim$ 40 eV)
at the emission lines of C–K, O–K, Cu–L, Al–K, Ti–K, Fe–K, Cu–K, and Ge–K. 
The charge transfer inefficiency (CTI) was found to be extremely 
low (e.g., only $5.9 \times 10^{-5}$ at C–K for TM7) yielding excellent spectral resolution for a CCD (e.g., 49\,eV FWHM at
C–K, see Fig~\ref{Boron}),
with a typical 1$\sigma$ uncertainty of only $\pm$1\,eV in the absolute energy scale over the whole 0.3--10\,keV bandpass. The temperature dependence of the gain and CTI was found to be small, and no obvious deviations from spatial homogeneity in the sensitivity were seen.
No evidence was found for any leakage of X-rays generated by the $^{55}$Fe calibration source. The energy resolution of the cameras are listed in Table \ref{ergres}.

The quantum efficiency (QE) of the cameras has been compiled from separate measurements of the pnCCD detector\footnote{data taken from: \\https://nbn-resolving.org/urn:nbn:de:hbz:467-6559} and of the various filters (Al, Al\,+\,PI), all of which were carried out at the BESSY synchrotron facility.

\subsection{Mirror calibration}
\begin{figure}[h]
\includegraphics[width=9cm]{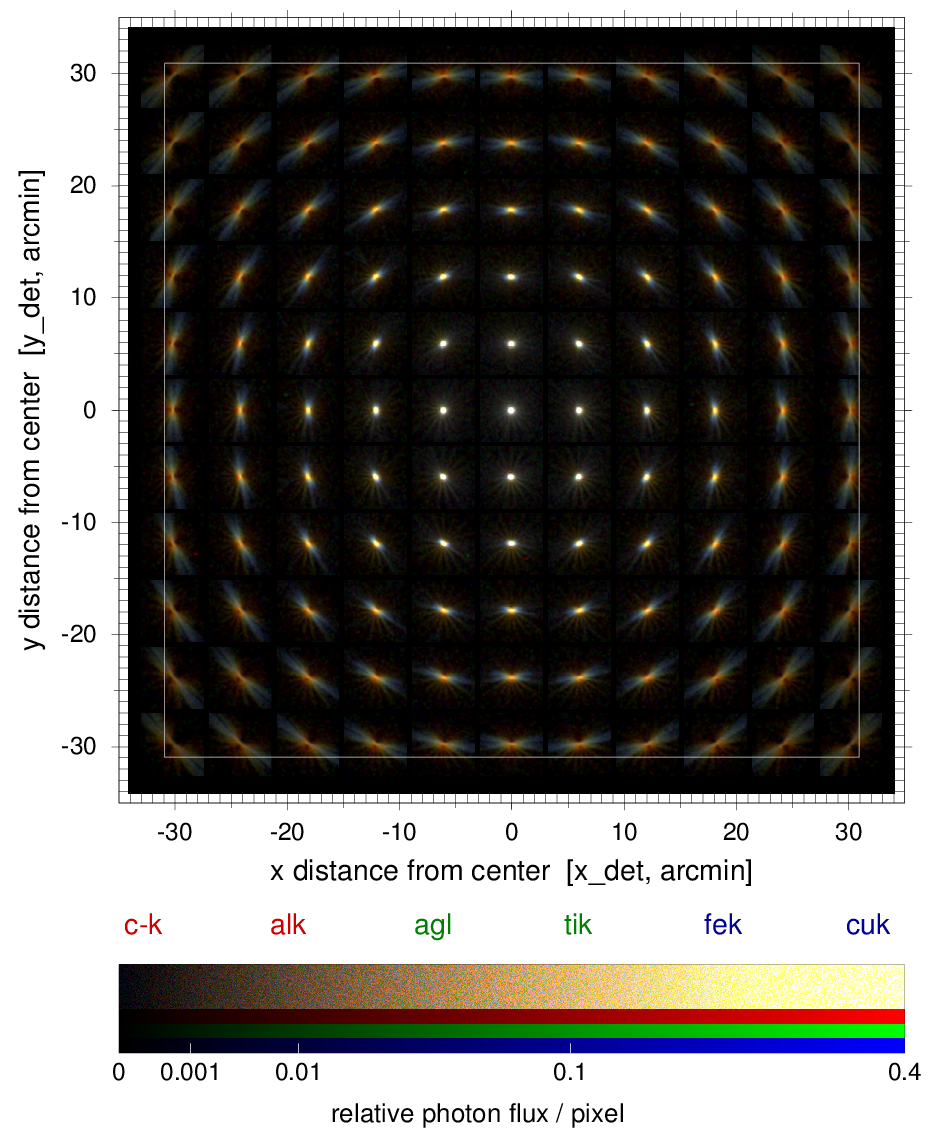}
\caption{
Visualization of the PSF mapping (TM1). The measurements were taken at various energies and the RGB PSF 
images created by combining the following datasets: C-K and Al-K (red), Ag-L and Ti-K (green), Fe-K and Cu-K (blue). 
The overall brightness is proportional to the relative photon flux per pixel.
}
\label{blurring}
\end{figure}

\begin{figure}[h]
\includegraphics[width=9cm]{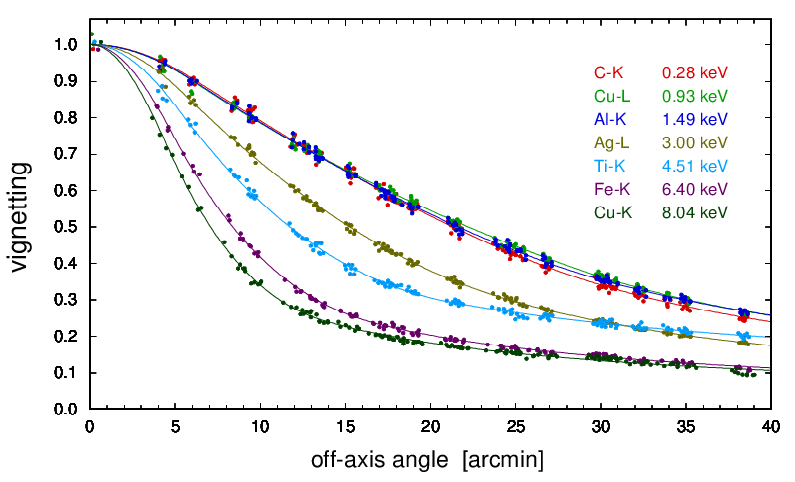}
\caption{Vignetting curves at seven different energies for TM1. The dots show for point sources the relative encircled flux within r = 4.0 arcmin derived from 1073 PSF images covering the focal plane (see Fig. \ref{blurring}); the curves show the derived parametrized vignetting function. Curves for the other TMs are similar.}
\label{vignetting}
\end{figure}

The calibration measurements of the eROSITA MAs started in
August 2014 and lasted until June 2016. They were all performed at the
PANTER 130\,m long-beam X-ray test facility of MPE
\citep{Burwitz2014}\footnote{According to the test plan it was foreseen to
use a flight-like camera ``eROqm'' for all calibration measurements. Due to
a failure of this camera all measurements were actually performed with
``TRoPIC'', a $256\times256$ pnCCD camera that had already been used for
most of the acceptance test measurements.}. The main goals of the
calibration were to predict the optical performance of each MA in orbit and to provide data for the correct mounting of MAs and CAs
in the telescope structure \citep{Dennerl2012}. For example, precise
knowledge of the focal distance is necessary for the correct relative
positioning of camera and mirror: In flight configuration, the cameras are
mounted in an intrafocal position (by 0.4~mm),  which leads to a slight
degradation of the on-axis performance compared to the values quoted
here (to about $18''$ on average), but improved angular resolution averaged
over the field of view.

The on-axis PSF was calibrated at three energies (C-K, Al-K, Cu-K) in
standard configuration with the X-ray source distance at 124\,m. 
The effective area determination at eight energies (C-K, Cu-L, Al-K, Ti-K, Cr-K,
Fe-K, Cu-K, Ge-K) consisted of several measurements of small aperture segments, which were illuminated one after the other with a quasi-parallel beam thereby
overcoming the problem of the finite source illumination. The off-axis
effective area (vignetting) was calibrated in an 1 degree $\times$ 1 degree
field using the focal plane mapping shown in Figure~\ref{blurring}.
However, the full set of calibration measurements was 
only performed on TM7 (for PSF and on-axis effective area) and TM1 (for
the off-axis PSF and vignetting).
Both PSF and effective area measurements were analyzed within an ``acceptance circle'' with 7.5mm (16 arcmin) radius.

\begin{table*}
\caption{Some key performance parameters of the eROSITA MAs as calibrated on ground: The on-axis angular resolution (HEW [arcsec]) is corrected for the detector resolution. The PSF has been measured also at C-K but is omitted here because it is almost identical to Al-K. The FWHM [arcsec] is an approximate value of the mirror--detector combination. The on-axis effective areas [cm$^{2}$] were measured using the standard setup. Errors are 1$\sigma$ for PSF and 3$\sigma$ for effective areas.}
\label{psfs}      
\centering          
\begin{tabular}{l c c c c c c c c }     
                                           & TM1          & TM2          & TM3          & TM4          & TM5          & TM6          & TM7  \\
\hline                   
     HEW Al-K$\alpha$ at 1.49\,keV          & 16.0$\pm$0.2 & 15.5$\pm$0.2 & 16.5$\pm$0.2 & 15.9$\pm$0.2 & 15.5$\pm$0.2 & 15.6$\pm$0.2 & 17.0$\pm$0.2 \\ 
     FWHM Al-K$\alpha$ at 1.49\,keV         & $\sim$9.3    & $\sim$7.0    & $\sim$7.9    & $\sim$7.6    & $\sim$8.5    & $\sim$7.9    & $\sim$9.2    \\
     HEW Cu-K$\alpha$ at 8.04\,keV          & 14.5$\pm$0.2 & 15.1$\pm$0.2 & 15.6$\pm$0.2 & 16.3$\pm$0.2 & 15.1$\pm$0.2 & 16.2$\pm$0.2 & 14.7$\pm$0.2 \\
     FWHM Cu-K$\alpha$ at 8.04\,keV         & $\sim$7.9    & $\sim$7.5    & $\sim$6.5    & $\sim$7.6    & $\sim$6.6    & $\sim$7.8    & $\sim$5.7    \\
     Eff. Area at Al-K$\alpha$ at 1.49\,keV  & 391$\pm$22   & 393$\pm$16   & 388$\pm$19   & 369$\pm$25   & 378$\pm$19   & 392$\pm$25   & 392$\pm$16   \\
     Eff. Area at Cu-K$\alpha$ at 8.04\,keV  & 24.9$\pm$1.1 & 25.1$\pm$1.2 & 24.1$\pm$0.6 & 23.8$\pm$0.9 & 25.1$\pm$1.1 & 25.0$\pm$0.9 & 24.8$\pm$0.8 \\
\hline                   
\end{tabular}
\end{table*}

\begin{table*}
\caption{On-axis effective area of TM7 with 3$\sigma$ errors. These measurements were subdivided into several small aperture segments, which each are illuminated one after the other with a quasi-parallel beam thereby overcoming the problem of the finite source distance.}
\label{effar}      
\centering          
\begin{tabular}{l c c c c c c c c c}     
\hline  
    X-ray source           & C-K             & Cu-L          & Al-K          & Ti-K         & Cr-K         & Fe-K          & Cu-K          & Ge-K         \\
    Energy [keV]           & 0.28            & 0.93          & 1.49          & 4.51         & 5.41         & 6.40          & 8.04          & 9.89         \\
        eff. area [cm$^{2}$]   & 356.2 $\pm$12.0  & 330.1$\pm$7.5 & 367.9$\pm$6.9 & 75.3$\pm$6.0 & 52.8$\pm$3.3 & 36.1$\pm$ 0.9 & 19.41$\pm$0.6 & 7.71$\pm$0.3 \\
\hline                   
\end{tabular}
\end{table*}

\subsection{Telescope performance}

\begin{figure}[h]
\includegraphics[width=9cm]{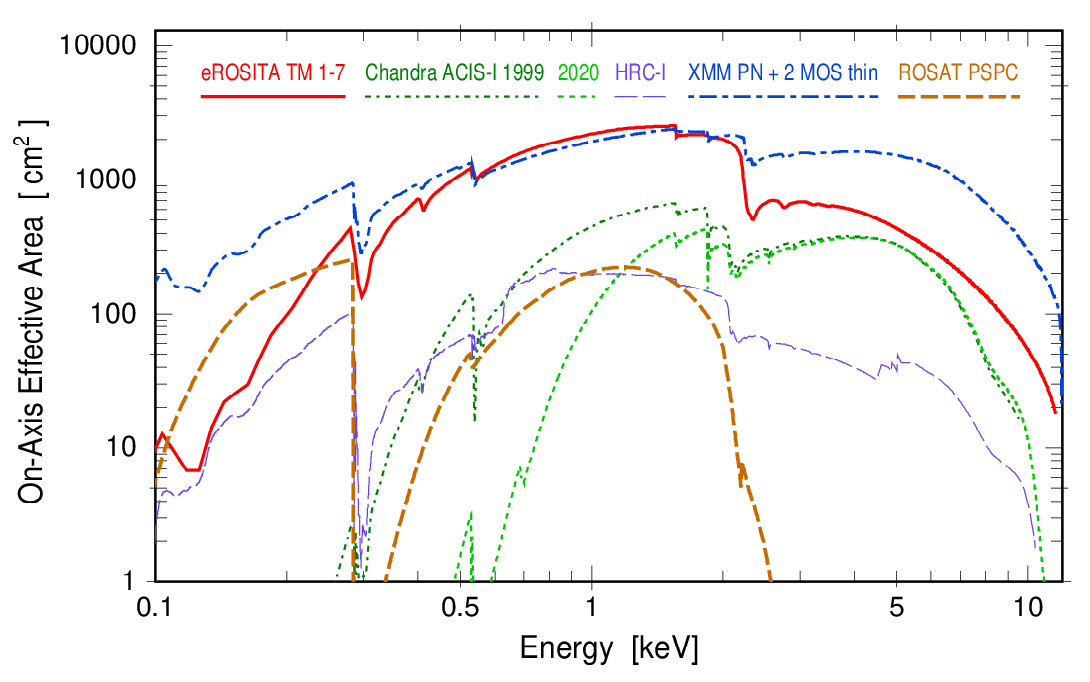}
\caption{Comparison of the on--axis effective areas as a function of energy for eROSITA (red), \chandra ACIS-I (in 1999, dark green, and in 2020, light green), \chandra HRC-I (purple), \xmm (blue), and ROSAT (brown).}
\label{effarea}
\end{figure}

\begin{figure}[h]
\includegraphics[width=9cm]{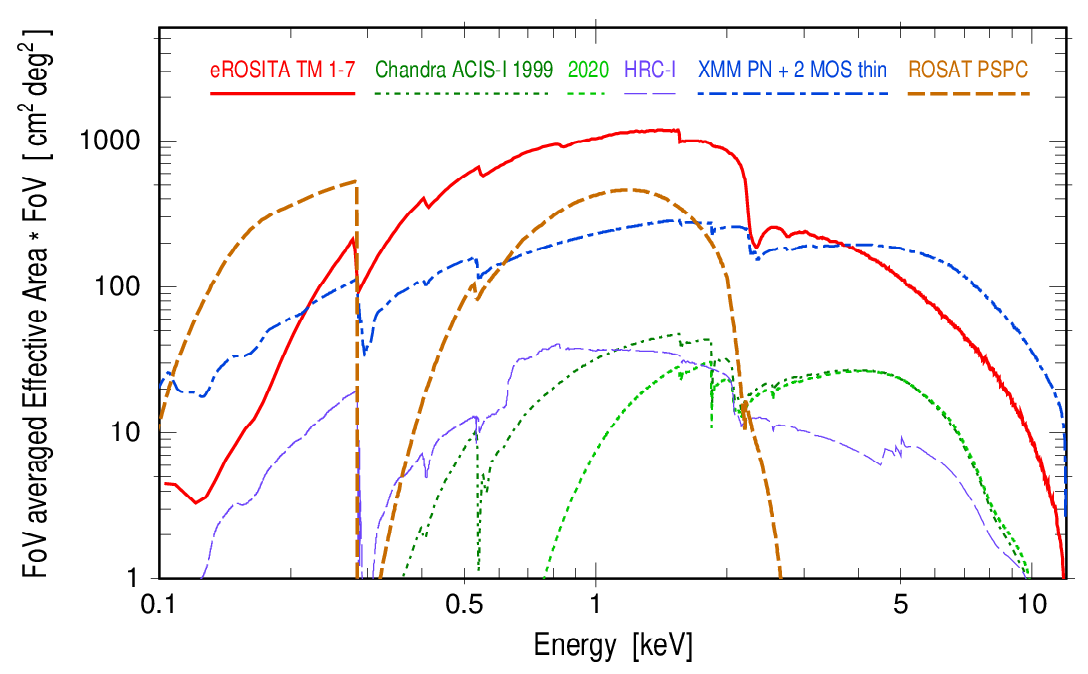}
\caption{Comparison of the grasp, defined as the product of field of view multiplied by (averaged) effective area as a function of energy for eROSITA (red), \chandra ACIS-I (in 1999, dark green, and in 2020, light green), \chandra HRC-I (purple), \xmm (blue), and ROSAT (brown).}
\label{grasp}
\end{figure}

The calibration campaign of the telescope (all mirrors and all cameras individually) demonstrated the excellent performance of the instrument (Tables~\ref{ergres},~\ref{psfs}, and~\ref{effar}), 
with the key parameters close to or exceeding the specifications defined in the early phases of the mission \citep{Predehl2006SPIE}.

The combined effective area (on--axis) of the seven eROSITA telescopes is slightly higher that of \xmm pn $+$ MOS (Fig.~\ref{effarea}) in the key 0.5-2 keV band. The ``grasp'' of eROSITA, defined as the product of field of view muliplied by (average) effective area, is shown in Fig.~\ref{grasp} in comparison with  \xmm pn $+$ MOS, \chandra,  and ROSAT PSPC \footnote{Chandra data are taken from: \\  
https://cxc.harvard.edu/cgi-bin/prop\_viewer/build\_viewer.cgi?ea, 
XMM-Newton and ROSAT data from: \\
https://heasarc.gsfc.nasa.gov/cgi-bin/Tools/w3pimms/w3pimms.pl}.

The grasp of eROSITA is the largest of all these imaging X-ray telescopes in the energy range $\approx 0.3-3.5$ keV
and clearly
highlights the major breakthrough it provides in terms of survey speed and wide-field imaging capability over a broad energy range.

\section{Background}
\subsection{Prelaunch estimates}

Prior to launch, the expected eROSITA background was simulated based on photon and high-energy particle
spectral components \citep[see e.g.,][]{Merloni2012}. 
The cosmic diffuse photon X-ray background has been adopted from the 
measurements with the \xmm EPIC cameras, as reported in \citep{Lumb2002} .
The high-energy particle background was calculated with Geant4 simulations \citep[see][]{Tenzer2010,Perinati2012}, 
with the mass model of eROSITA available at that time.

\subsection{In-orbit measurements}

\begin{figure}
\includegraphics[width=9cm]{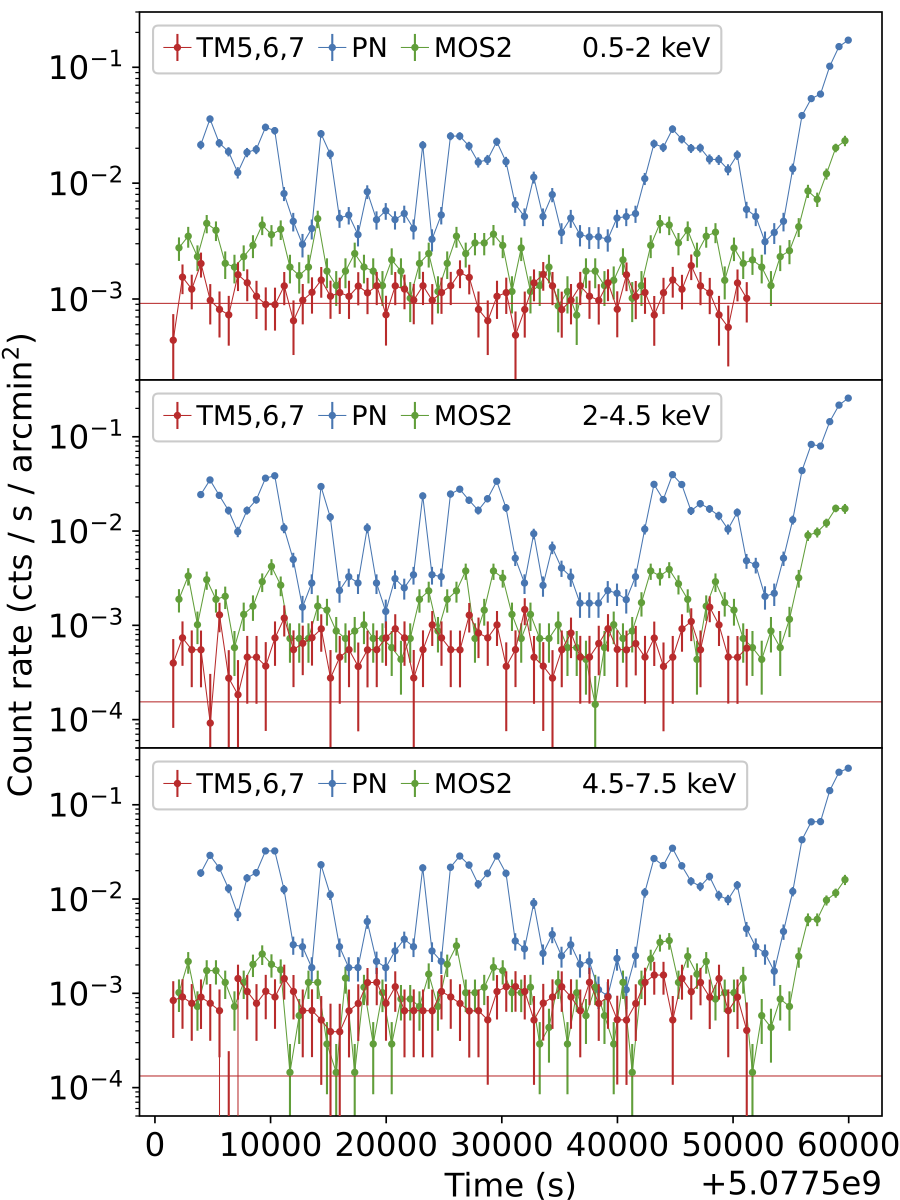}
\caption{Comparison of background light curves in different energy bands between eROSITA (TM5,6,7 only; red), XMM-Newton EPIC-pn (blue), and XMM-Newton EPIC-MOS2 (green) detectors. The data were taken simultaneously during the observation of the NLS1 1H0707-495 \citep{Boller2020}
on October 11, 2019, for 60 ks, while SRG was about to be inserted into its final halo orbit around L2. Top panel: 0.5-2\,keV; Middle panel: 2-4.5\,keV; Bottom panel: 4.5-7.5\,keV. The thin red lines in each panel show the predicted background level from pre-launch estimates \citep{Merloni2012}, rescaled by a factor of $3/7$ to account for the fact that only three eROSITA
TMs were operational during this observation. }

\label{BkgCalPVlightcurve}
\end{figure}

After launch, during the eROSITA commissioning, a higher particle background than expected from pre-launch estimates from 2012 was observed.
Figure~\ref{BkgCalPVlightcurve} shows a comparison of the background count rates measured over three broad energy bands simultaneously by eROSITA and XMM-Newton. Compared to the pre-launch estimates (thin red lines in Fig.~\ref{BkgCalPVlightcurve}), the eROSITA background is consistent at energies below 2\,keV, but a factor of a few higher at energies above 2\,keV, as would be expected if the level of (un-vignetted) particle background were higher than predicted. On the other hand, the eROSITA background shows a much higher degree of (temporal) stability  compared to that measured by XMM-Newton\footnote{Rationale for comparing eROSITA with XMM-Newton in terms of
``cts/s/keV/arcmin$^2$'':
The eROSITA CCDs are operated in framestore mode.
The framestore area is shielded against celestial X-rays but is sensitive to instrumental background caused by high-energy particles.
Those background events get also coordinates assigned to the imaging area -- and are then also projected onto the sky.
This projected background is the relevant quantity for source detection of faint and slightly extended objects like clusters of galaxies.}, where the well-known large background flares due to soft protons are evident.

As the overall background at high energies is dominated by the particles interacting with the instrument (and instrument noise), the spectrum of all events measured by eROSITA above 2\,keV combining data taken from the first all-sky survey (Fig.~\ref{BkgSurveyChurazov}) is consistent (in shape) with the background spectrum as measured with the filter wheel setting to a closed position (Fig.\ref{BkgSurveyClosed}).  

\subsection{Comparison with pre-launch expectations}

\begin{figure}
\includegraphics[width=9cm]{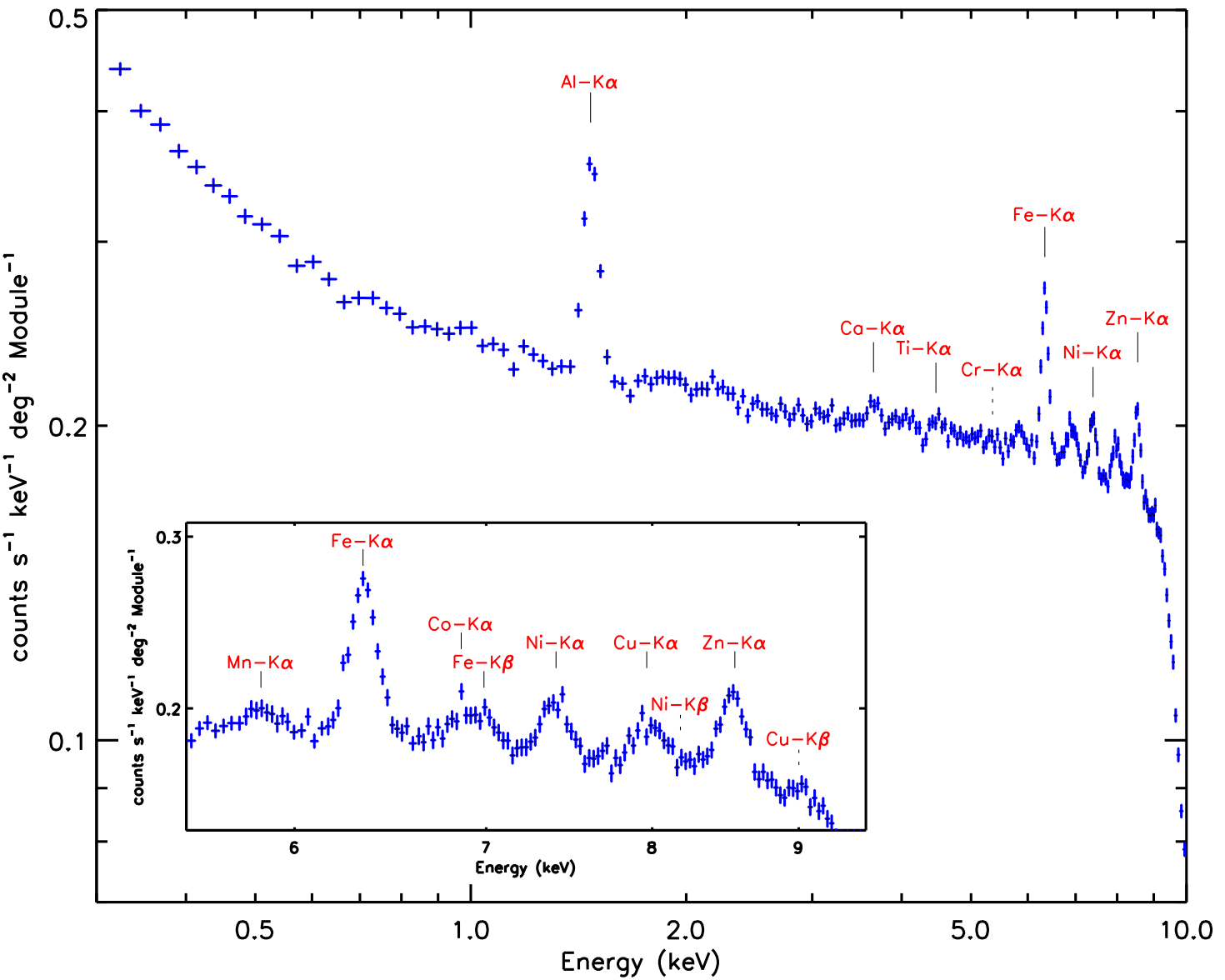}
\caption{Combined eROSITA camera background spectrum measured in orbit in CLOSED setup for the five modules with on-chip filter (TM\,1,2,3,4,6). The effective exposure time is \mbox{514\,ks} for one single module. Red markers indicate the nature of the most significant fluorescence lines.}
\label{BkgSurveyClosed}
\end{figure}

\begin{figure}[h]
\includegraphics[width=9cm]{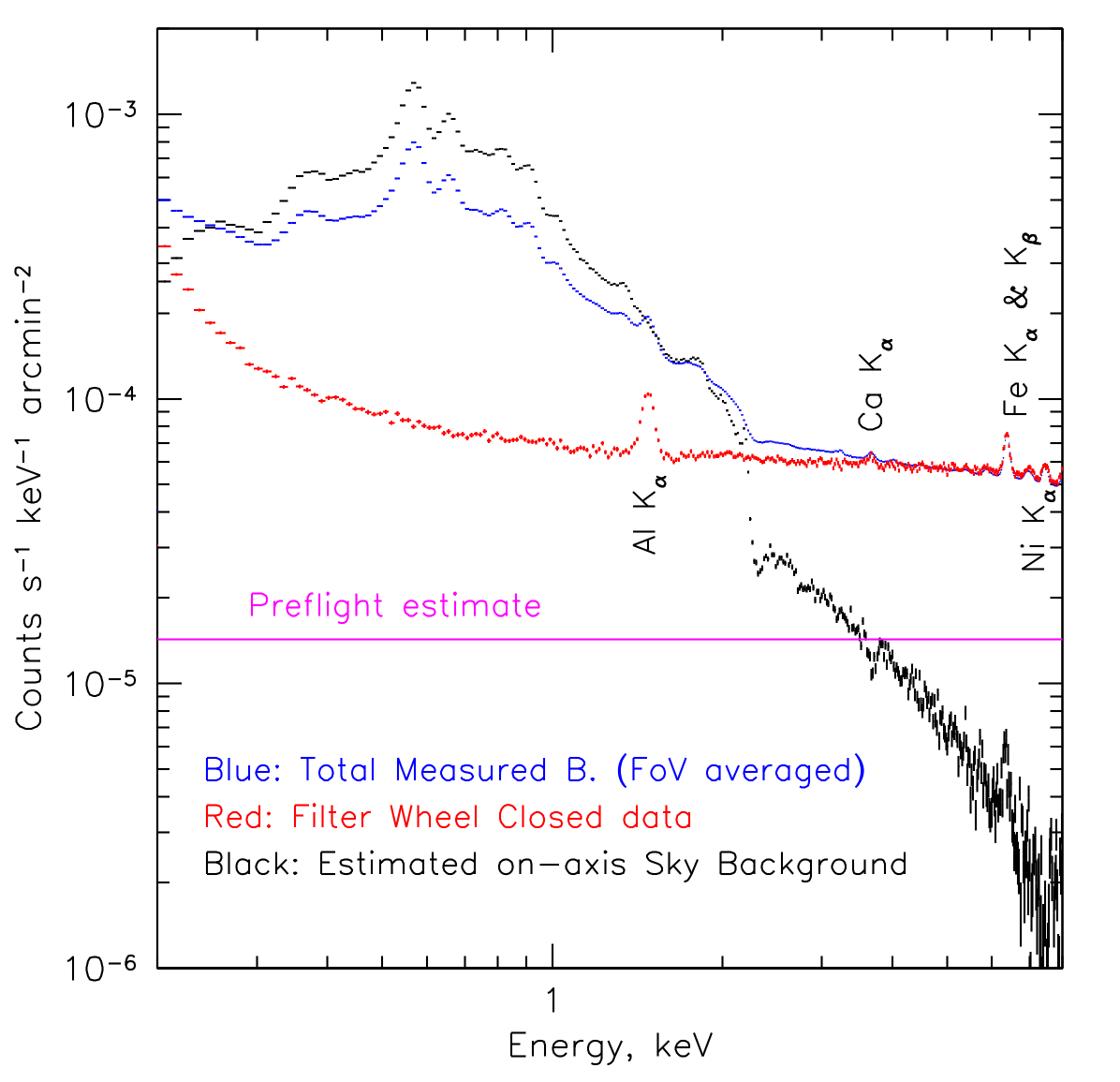}
\caption{Stacked eROSITA all-sky survey background spectrum (per individual telescope). The blue points are the total measured background. The red points show the background measured during filter wheel closed observations (see Fig.~\ref{BkgSurveyClosed}). The indicated spectral lines are of instrumental origin (see text).
The black points mark the reconstructed 
(i.e., corrected for vignetting) on-axis photon background spectrum. The horizontal magenta line is the approximate expected particle background level estimated pre-launch \citep{Tenzer2010}. }
\label{BkgSurveyChurazov}
\end{figure}

Despite the presence of the dedicated graded shield, line features are seen in the 6--9\,keV range which are presumably excited inside the camera as the closed filter wheel position and the proton shield (3\,cm copper)
effectively block such X-rays from the outside (Fig.\ref{BkgSurveyClosed}). 
The spatial distribution of these lines appears to be very homogeneous,
unlike in the case of EPIC-pn aboard XMM-Newton. This indicates that the beryllium of the graded shield itself might be the origin due to impurities. Preliminary simulations based on an analysis by \citet{McGarry} support this, but require further laboratory tests. In contrast, the strong Al-K$\alpha$ line is a result of the closed filter wheel, that is, 4\,mm of aluminium in the optical path between mirror modules and CCD (Fig.\ref{BkgSurveyClosed}).

Several factors could contribute to the higher instrumental background measured by eROSITA, compared to pre-launch expectations. 
The Sun is currently at a minimum in terms of activity, which results in the highest Galactic cosmic ray flux incident on the instruments in space. The pre-launch predictions on the other hand assumed a launch date close to solar maximum, and hence with the lowest incident cosmic ray flux. 

The anti-correlation of observed instrumental background with the solar cycle is also known from XMM-Newton and Chandra observations (e.g., \citealt{XMM-SOC-GEN-TN-0014}; \citealt{Grant2014SPIE}). Also, the mass model used in the early predictions did not include all camera components in detail. This could have led to an 
underestimate of the background from higher-$Z$ materials (with higher fluorescence yields)
 present in smaller parts of the structure, or as ``contamination'' in low-$Z$ materials (see above).
We are currently revisiting both the eROSITA instrument Geant4 physical model based on the final flight model design, and the models for incident particle spectra. A detailed analysis of the measured particle background and the comparison with the new model predictions will be presented elsewhere (Eraerds et al., in prep.).
It should be noted that this is the first time that the X-ray background in all its components has been measured in L2.

\section{Ground software and data analysis pipeline}

A ground software system was developed in parallel to the instrument hardware  \citep{Brunner2018}. Building on the experience and in part on code from the \xmm and ROSAT X-ray observatories, the eROSITA Science Analysis Software System (eSASS) provides an environment and tools for the creation of calibrated science data products and to perform various interactive data analysis tasks. 

\begin{figure}
\includegraphics[width=9cm]{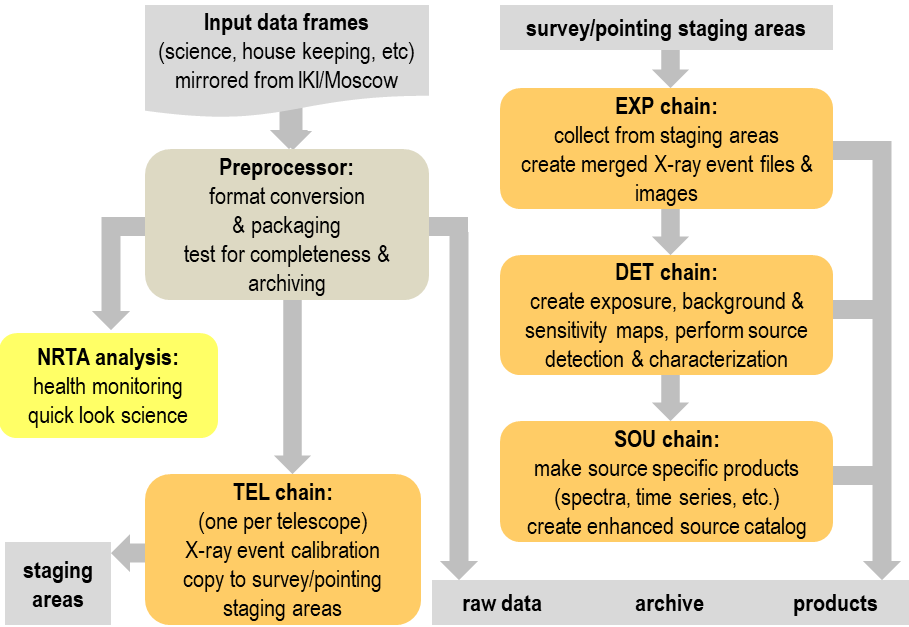}
\caption{\textbf{eSASS data analysis pipeline}, consisting of four task chains creating a full set of calibrated data products. This pipeline is fed by a pre-processor which also provides input to the near real-time analysis.}
\label{PipeDiagram}
\end{figure}

\subsection{Telemetry and pipeline}

The telemetry data from eROSITA received during each daily SRG ground contact
are converted into the standard FITS format by a pre-processor which
re-formats, packages, and archives the data received during
each data dump. Once the data for a given time interval are acquired,
they are processed using a software pipeline to create a set of calibrated data
products, including calibrated X-ray event lists, images, exposure,
background, and sensitivity maps, as well as various X-ray source
catalogs and associated source and calibration data products. The data
processing pipeline is organized into task chains for event
calibration, exposure creation, source detection, and extraction
of source-specific data products. The layout of the data analysis
pipeline is shown in Fig.~\ref{PipeDiagram}.

The pipeline supports all three main eROSITA observing
modes (all-sky survey, pointing, field scan). The all-sky survey data
products are organized into 4700 overlapping sky tiles of
$3.6^\circ \times 3.6^\circ$ in  size, which are updated on a daily basis as new data are received. Data products are provided individually for each six-month full-sky coverage as well as cumulatively for each survey update. 

All pipeline data products are made available to authorized users
through a web interface which supports data requests by observation,
sky field ID, or region. Astrometrically
corrected catalogs of detected X-ray sources are updated and made available to the consortium weekly.

A total of 1,004,624 X-ray sources were detected in the first of eight all-sky surveys on both hemispheres (see section ~\ref{sec:erass} below).
Daily updated all-sky maps tracking the progression of the four-year all-sky survey are accessible to members of the science teams via a web interface. 

\subsection{eSASS and NRTA software}

For in-depth interactive data analysis, the eSASS package provides command-line tools performing such 
functions as data selection and binning, source detection and characterization, and the creation of spectra 
and light curves. Details of the tools and algorithms are described in more detail elsewhere (Brunner et al. 
2020, in preparation). All data products are FITS files that are largely compliant with FITS standards, meaning  that a range of popular X-ray data analysis tools may be used in addition to the eSASS.

The eSASS software package interacts with a calibration database maintained by the eROSITA calibration team, which provides calibration products for telescope alignment (boresight), mirror vignetting, and point response functions, energy calibration, detector response, and effective areas, among others.
 
The standard processing performed by the eSASS is complemented by a near
real-time analysis (NRTA) which provides interactive instrument health
checking as well as quick-look science analyses
\citep{kreykenbohm:09a}. The purpose of this quick-look science
analysis is 
twofold. First, it enables rapid analysis of the data to identify anomalies, and second it allows the rapid identification of transient or strongly variable
sources. The NRTA includes a backend in which a set of several thousand known
sources are automatically monitored. 
Operators are alerted based on predefined trigger criteria
for these sources, which include eROSITA count rates,
X-ray colors, and the source history in eROSITA. The NRTA
provides a flexible way to define triggers based on combinations
of these parameters, allowing, for example, to trigger on unexpected source
hardening due to absorption events, the detection of a transient
source, or sources that are rapidly brightening in flares. In
addition, for selected source regions such as the Magellanic Clouds,
the NRTA also extracts preliminary source images based on the
downloaded telemetry, and performs a source detection to alert for new
X-ray sources through correlating the sources found in the downlinked
data with existing X-ray catalogs. Finally, in order to find bright
transient sources, NRTA also performs a Bayesian block search for
rapid count-rate changes \citep{scargle:98a}. This backend is
complemented with a web-based interface for viewing of housekeeping
and science data and for setting alert conditions for instrument
health monitoring and to support quick reaction to transient X-ray
events.


\section{Mission planning}
\label{sec:planning}

\begin{figure*}[h]
\includegraphics[width=18cm]{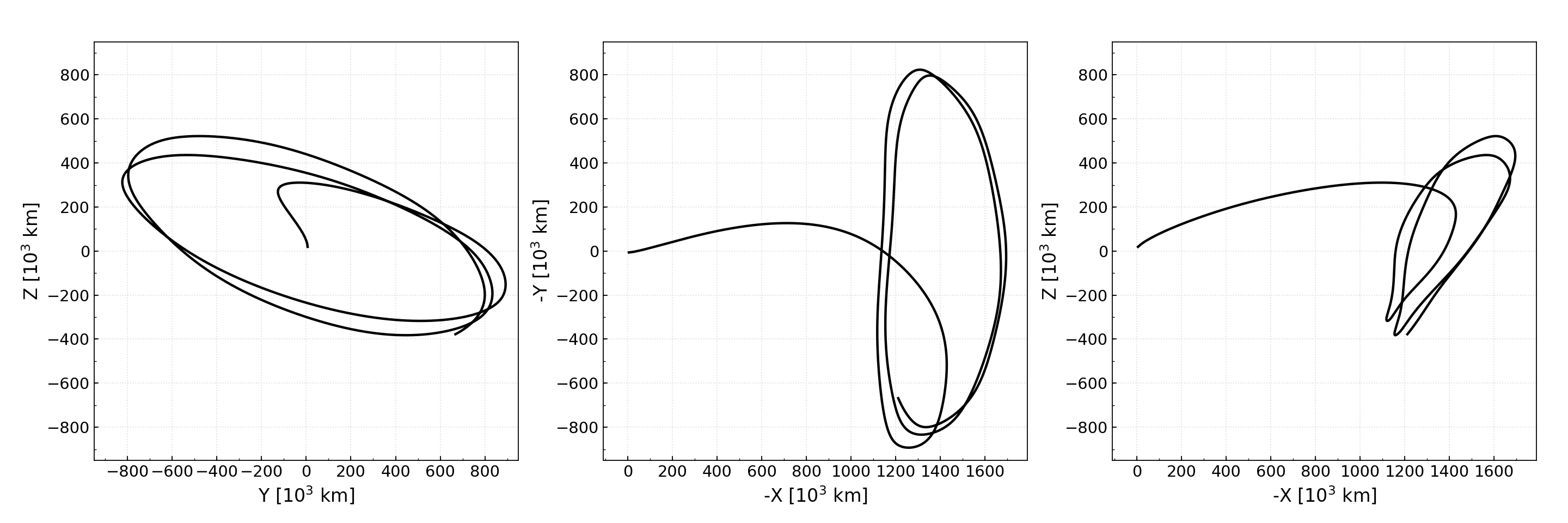}
\caption{SRG trajectory for the first 450 days after launch; image shows ecliptic projections of the L2 orbit in Geocentric Solar Ecliptic  
coordinates. The transfer to the L2 region took about 100 days, and the L2 orbit has a period of about half a year.}
\label{orbit}
\end{figure*}

The mission planning team interfaces the scientific and operational aspects of the eROSITA project. It prepares and schedules scientific observations, investigates strategies for optimal mission performance, and develops related software and data structures. The SRG mission planning is done by a joint German--Russian collaboration that includes eROSITA (at the Hamburger Sternwarte and at MPE) and ART-XC (IKI) team members, as well as spacecraft operations experts (NPOL).

\subsection{Mission panning boundary conditions}

The placement of SRG in an L2 orbit enables large operational flexibility, although typical angular orientation limits with respect to the Sun and Earth have to be considered. With the spacecraft axes being defined as $X_{\rm SC}$ (telescopes line of sight), $Y_{\rm SC}$ (solar panels extension direction), and $Z_{\rm SC}$ (rotation axis, antenna cone direction) in a right-handed coordinate system with $+X$ being the viewing direction and $+Z$ pointing to the inward Solar System direction (see also Figure \ref{fig:LLTM5}), the following angular constraints have to be taken into account:
\begin{itemize}
\item Sun angle: (a) The $Z_{\rm SC}$ axis has to be within $\pm 20^{\circ}$ of the Sun (solar panels operations, stray light mitigation, cooling balance) and (b) the angle between Sun-direction and the XOZ-plane has to be within $\pm 13^{\circ}$. As a consequence, during survey mode $Z_{\rm SC}$ has to be within $\pm 13^{\circ}$ of the Sun.

\item Earth angle: The $Z_{\rm SC}$ axis has to be within $\pm 24^{\circ}$ of the Earth during ground contact (antenna cone width). Keeping the Earth permanently within the cone of the onboard antenna allows for continuous survey operation.
\end{itemize}
The movement of SRG around the L2 point (see Fig. \ref{orbit}), with a period of about half a year, and the movement of the Earth around the Sun result in time-variable orientation restrictions for survey mode operations or corresponding observing windows for astronomical targets\footnote{A web-based interactive visibility calculator tool is available at \url{erosita.hs.uni-hamburg.de}}.
In addition, the generated mission time-line has to comply with ground contact intervals, orbit correction maneuvers, and other technical operations.

\subsection{Observing modes}

SRG/eROSITA can be operated in three observing modes, namely: survey, pointing, and field scan. All modes have been tested successfully and can be interleaved.
In survey mode the spacecraft rotates continuously, thereby scanning great circles on the sky. This is the prime observational mode during the all-sky survey phase.
In the pointing mode a single target or sky position is observed for a given time, while in field scan mode a sky region of up to $12\fdg{}5\times12\fdg{}5$ in size is scanned in a rectangular grid pattern. Each field scan is adapted to the respective scientific requirements, making it an important capability of SRG.
During the calibration and performance verification phase, over 100 individual pointings and field scans were performed with eROSITA as prime instrument between mid-September and mid-December 2019.

\subsection{The all-sky survey}
\label{sec:erass}
\begin{figure*}[t]
\includegraphics[width=18.4cm]{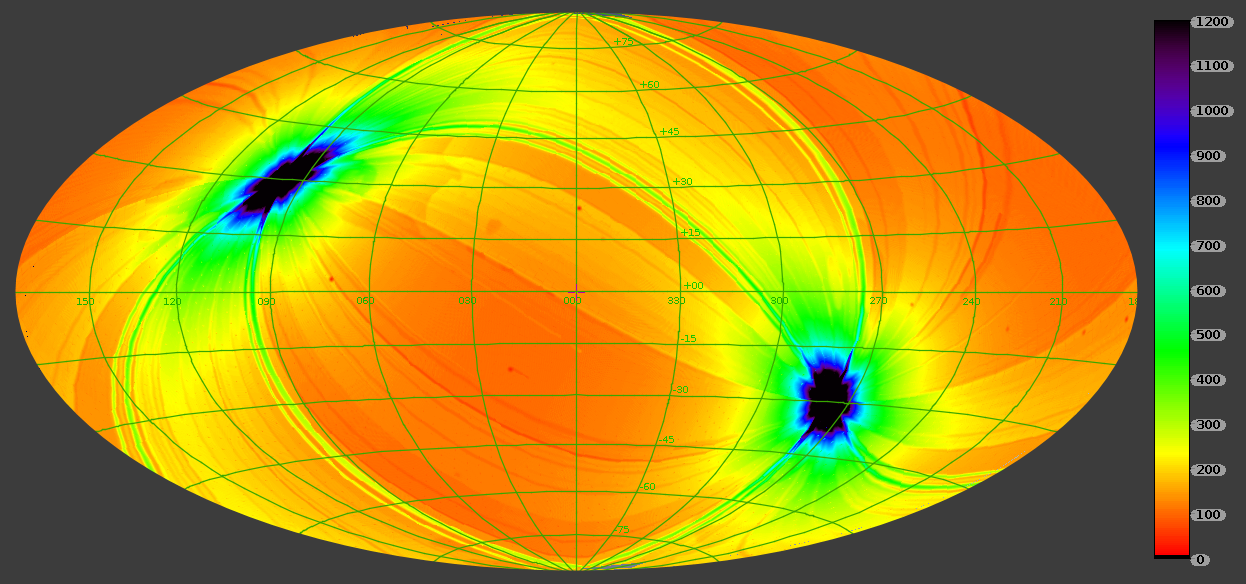}
\caption{
Effective (vignetted) exposure map derived from the first all-sky survey, eRASS:1 (galactic coordinates, Aitoff projection). The values in the map show the exposure time multiplied by the average of the ratio of the (vignetted) effective area to the on-axis effective area in the energy band 0.6-2.3\,keV). Uncorrected (nonvignetted) exposure times are about a factor of 1.88 higher in this energy band.
Effective exposure values range from ${\sim}100$\,s at the ecliptic equator to more than $10000$\,s close to the ecliptic poles.} 
\label{SurveySEP}
\end{figure*}

The implementation of the eROSITA all-sky survey (eRASS) is defined by the survey strategy with three basic parameters.
First, the ``scan rate'' defines the rotation of the spacecraft. Here an
angular velocity of $0.025\,\mathrm{deg}\,\mathrm{s}^{-1}$ is used, a
spacecraft revolution has a duration of 4\,hr and a central field-of-view (FOV) passage
time of about 40\,s. This rate avoids a deterioration of the angular resolution of the instrument and provides sufficient overlap between subsequent scans.
Second, the `survey rate' describes the progression of the scanned great circles on the sky. With an average angular velocity around the Sun of about $1\,\mathrm{deg}\,\mathrm{d}^{-1}$, scans are separated by about $10'$ each and any position on the current scan path is observed roughly six times per day.
Third, the `survey pole' defines the plane in which the rotation
axis is moving; its choice primarily influences the exposure at the
polar regions,that is, the location, extent, and depth of the deeper
exposed regions. This can be used, for example, to create larger
regions with deep exposures below source-confusion limits or to
increase the exposure depth at sky regions of specific interest. The
eRASS started with the ecliptic pole as survey pole. Figure~\ref{SurveySEP} shows the final effective exposure map in the 0.6-2.3\,keV band of the first all-sky survey, completed between December 13, 2019, and June 12, 2020. A modification of the survey pole after some time is envisaged.

The survey geometry approximately follows ecliptic coordinates. This results in a latitudinal exposure distribution with lowest exposure values close to the ecliptic plane and highest exposures at its poles where the scan paths overlap.
An additional longitudinal exposure pattern is generated by a nonuniform angular movement of the spacecraft rotation axis, which is required in continuous survey operation to compensate angular separations between spacecraft--Sun and spacecraft--Earth directions larger than the antenna cone. This counteracting ``swing-movement'' with respect to solar orientation leads to a fluctuating survey rate with a roughly 90 period. The longitudinal distribution largely persists in the all-sky survey, as the period of the L2 orbit is very similar to the duration of an individual all-sky scan. The average total (unvignetted) exposure in the ecliptic plane region after four years of the survey is approximately 1.6\,ks, whereas the all-sky average is $\sim$ 2.5\,ks; effective (vignetted) exposure values are a factor of 1.88 and 3.31
smaller, for 0.2-2.3\,keV and 2.3-8\,keV, respectively.

\begin{figure*}
\centering\includegraphics [width=18cm]{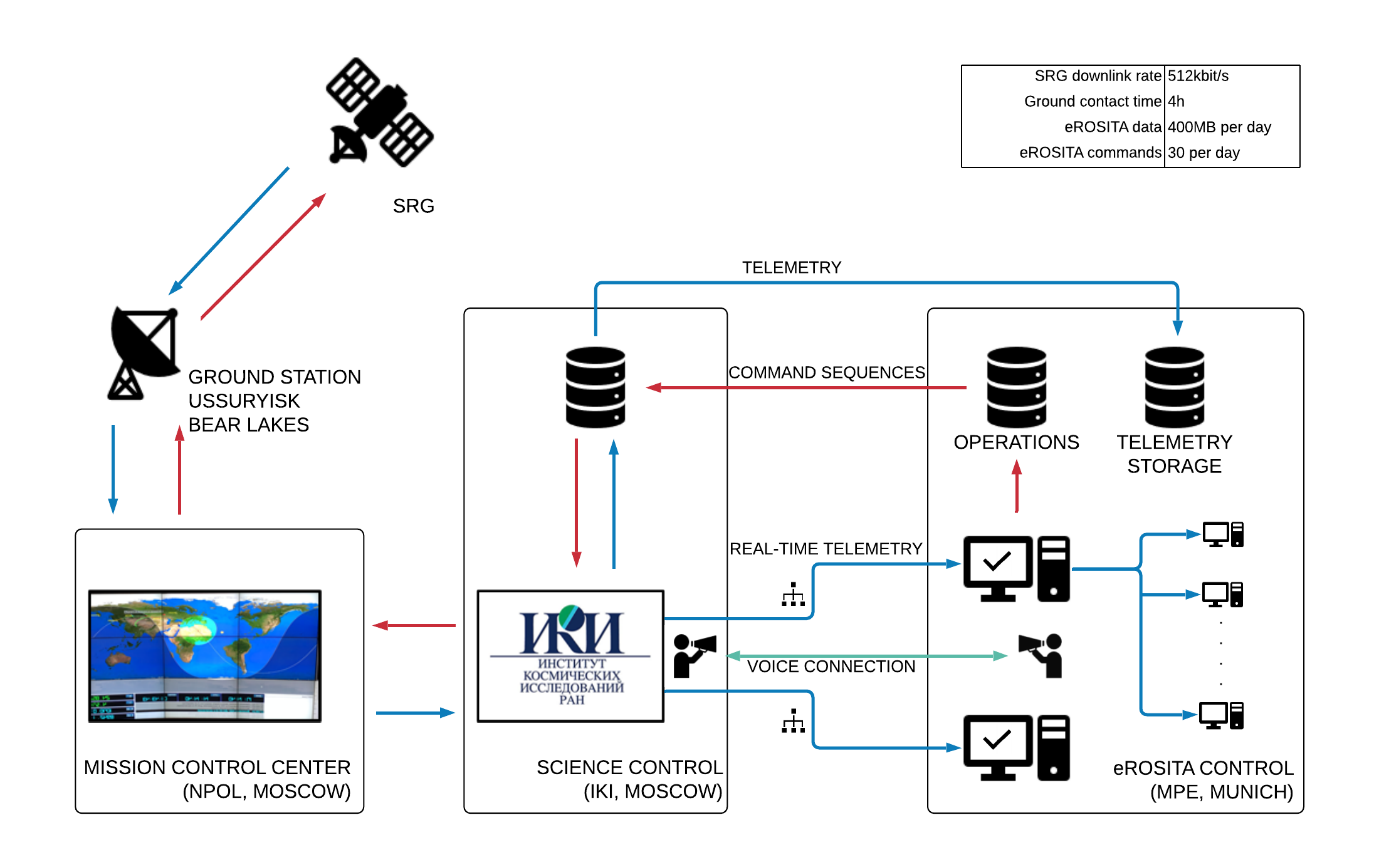}
\caption{eROSITA operations and ground segment.}
\label{fig:ops}
\end{figure*}

\section{eROSITA operations}

The SRG mission control is located at the mission control center (MCC) on the premises of NPOL  in Khimki near Moscow, where spacecraft control, flight dynamics, and ground antenna interfacing take place. Ground contact passes are planned on a daily basis with an  average duration of $\sim$4 hr. The main ground stations involved in operations are located in Bear Lakes near Moscow and in Ussuriysk in the Russian far east, and are part of the Russian Deep Space Network. The SRG is also compatible with ESA’s ESTRACK network of deep space antennae.

The mission scientific operations are shared between IKI and MPE, with MPE having full responsibility and control over eROSITA. The tasks consist mainly of verifying the health of the instrument, preparing command sequences to perform either routine or corrective actions, and dumping the science data collected in the mass memory storage of  eROSITA to ground. Ground contacts require the daily presence of eROSITA operators interfacing via a voice connection to IKI operators to carry out the command sequences and route the real-time telemetry to MPE (see Figure~\ref{fig:ops}). On average, around 60\% of the 
daily ground contact time is devoted to eROSITA, typically consisting of four stages:
\begin{itemize}
\item Monitoring of the telescope infrastructure, including the thermal control systems, the power distribution systems, and the general status of the complete instrument. This requires a low-data-rate link of 32~kbit/s.
\item Camera checks, where all housekeeping parameters of the seven CAs are reviewed and real-time photons are seen for camera integrity assessment. This requires the full-data-rate link of 512~kbit/s. 
\item Real-time commanding session, where maintenance and corrective actions are carried out with real-time eROSITA response. These commanding sessions are complemented with scheduled commands to be executed outside of ground contacts.
\item Mass memory data dump, where the complete telemetry is dumped from eROSITA's mass memory. This is on average 400~MB per day and requires the 512~kbit/s data-rate link.
\end{itemize}

The complexity of these operations resides in the fact that eROSITA is composed of seven independent instruments (TMs and CAs), which require individual monitoring and maintenance. More than 1000 housekeeping parameters are monitored daily and independent command sequences are generated for each CA. More than 10000 commands have been up-linked to eROSITA and approximately 88~GB were dumped to ground over the first 9 months of space flight. Despite the complexity of the instrument, the proximity of instrument developers, operations team, and science ground segment make eROSITA operations at MPE the optimum solution for such an instrument.


\section{Commissioning}

The commissioning phase of eROSITA had the objective of switching on all subsystems, and verifying they were functional following the launch and that they performed as expected to fulfil the scientific objectives. Finally, the commissioning was used to set up the telescope and its seven cameras into operational mode to start the Cal-PV phase. The simplified steps to fulfil these objectives were the following:

(1) verification of the electrical systems, (2) optimization of thermal settings, (3) opening the cover and filter wheels,
(4) opening the valves to fill the VCHPs and cool down the CCD Cameras, 
switch on CCD cameras and check health, (5) functionality, and performance.
and (6) verify the software functionality of ITC and CEs.

This phase served not only to verify and commission the complete eROSITA telescope, but also gave the ground teams in Khimky and Moscow (NPOL, IKI) and Garching (MPE) the opportunity to learn and update the procedures on how to safely operate the spacecraft and the telescopes in space.

There were various constraints on the speed at which these activities could be performed. On the one hand, the ITC had to be switched on less than 4 hr after launch to enable the thermal control of mirrors and electronics. This was a mission-critical event, as the cooling of electronics through the radiators could quickly bring them to temperatures under -50$\degree$C and render them useless. On the other hand, cover opening, camera cooling, and camera switch-on had to wait several days before being activated to avoid excess contamination from the first two spacecraft burns taking place on day 10 and day 20. In addition, camera cooling could not be performed without a minimum of 21 days of outgassing following the cover opening. These constraints led to the commissioning sequence detailed in Table~\ref{tab:milestones}.

Despite the fulfilment of the commissioning phase, two major issues were highlighted during the electronics verification and camera commissioning: the first is related to a radiation susceptibility seen in the CE with respect to single event upsets (SEUs), and the second is related to a light leak detected in the camera platform that affects the cameras of TM5 and TM7. These issues are described hereafter, as they have an impact on the telescope operations.

\subsection{Camera electronics SEU susceptibility}
On August 10, 2019, during CE verification and commissioning, a processor crash was detected on one of the seven CE units, TM5. This crash had side effects related to the camera control voltages which were set without having been commanded. This initial CE disturbance together with two more disturbances on different CEs led to an interruption of the telescope commissioning. A delta-commissioning phase was defined, to further understand the problem and minimize the risk of any damage. The conclusions of this phase were the following:

\begin{itemize}
\item These disturbances can occur in any of the seven CEs.
\item They are digital disturbances that originate from different modules of the logic within the FPGA. 
\item The most probable origin of these disturbances are SEUs in the logic of the FPGA caused by high energetic Galactic cosmic rays.
\end{itemize}

As discussed above, the CE reads out and processes the CCD frames from the Camera and is based on a Virtex 4 FPGA carrying a PPC processor. That is the brain of the frame processing of each CE. Due to the complexity of the on-board frame processing, it was not possible to include triple module redundancy (TMR) in the CE FPGA, which is a technique used to prevent SEUs. This is the most probable reason behind the CE disturbances. These disturbances have continued to occur randomly on each CE throughout the mission so far, at a rate of approximately one disturbance per week for all cameras together. Each disturbance puts a camera out of order for an average of 12hr, after which it is reset and returns to normal operation.  

It is worth noting that the ITC also carries the Virtex 4 FPGA, but given the mission critical functionality of the ITC, full TMR is implemented there. In 10 months of continuous operation it only once had a digital disturbance that required a reset. This supports the claim that the CE susceptibility to cosmic radiation comes from lack of full TMR implementation on the FPGA.

Despite these disturbances, the observing efficiency of eROSITA, as measured after the first 6 months of scientific operations, is kept to more than 95\%.

\subsection{Light leak}
\begin{figure}
\centering\includegraphics [width = 9cm] {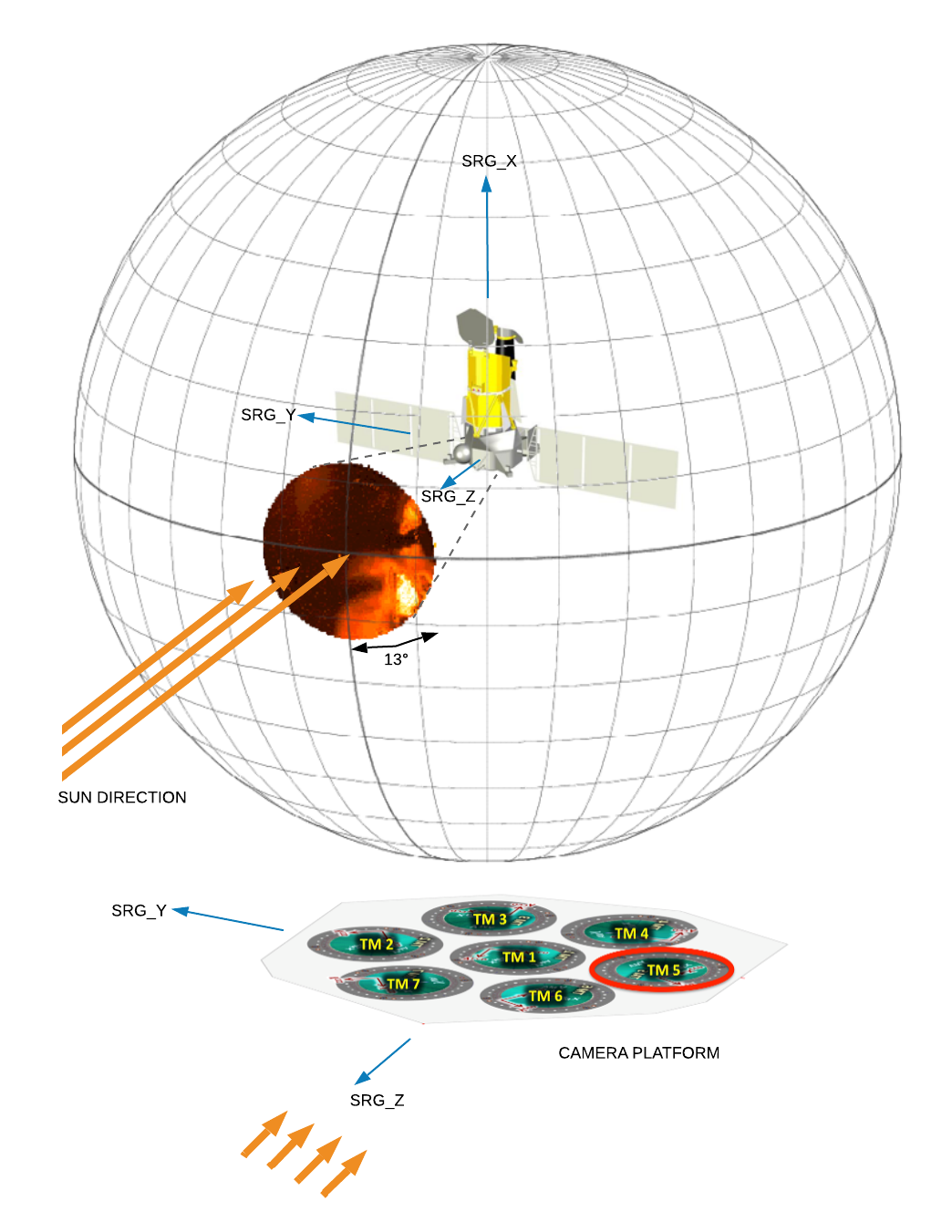}
\caption{Dependence of the optical light leak on the Sun direction: SRG is located at the center of a sphere, which illustrates the orientation of its principle axes (SRG$\_$X, SRG$\_$Y, SRG$\_$Z). 
The circular map on this sphere shows how the observed event rate in TM5 depends on the Sun direction. This map covers the full range of angles between Sun direction and SRG Z axis.  This map, which was derived from all-sky survey data, indicates that sunlight enters primarily from the lower right. The camera platform at the bottom shows where TM5 is located.}
\label{fig:LLTM5}
\end{figure}

During the commissioning of TM5 and TM7 it was noticed that optical light contaminated the bottom part of the respective CCDs.  It was later observed that the intensity of this illumination depended on the orientation of the telescope with respect to the Sun. The reason for only affecting these two cameras is the absence of the aluminium on-chip optical light filter that the other five cameras carry, as the plan was to use them for low-energy spectroscopy.
To limit the telemetry rate due to the optical contamination, the primary thresholds which were initially set to around 80 eV for all cameras, had to be increased by about 45-60 eV for TM5 and TM7. This had the consequence of decreasing the low-energy coverage and spectroscopic capabilities that were expected from these two cameras. At the time of writing, efforts are being made to understand where this light is coming from. Modeling of the occurrence and intensity of the light leak is also being pursued in an effort to minimize the loss of low-energy performance.
Indeed, during the first complete all-sky survey the characteristics of the light leak were better understood and it is now possible to describe them and propose mitigation actions that could potentially enable the operation of these cameras for what they were initially designed for, namely low-energy spectroscopy.
The mitigation actions are centered around the following aspects:
\begin{itemize}
\item The light leak is heavily dependent on the incident angle of the Sun on SRG. This changes throughout one complete survey because of the geometry of the orbit and the scanning plane. At certain angles the effects of the light leak are almost nonexistent on TM5 and TM7. These angles will be explored, and some new Sun angle constraints will be defined for analysis; see Fig.~\ref{fig:LLTM5}.
\item It was seen that there are periods of time within one SRG revolution that are almost unaffected by the light leak. These periods could be used to set the cameras to lower energy thresholds, therefore retaining the low-energy performance at least for some periods of time. This would enable spectroscopy of part of the sky at particularly low energies. The thresholds that could be used and the impact on the operational side will be analysed and a proposed approach will be defined.
\item The CCD image is not uniformly affected by the light leak on TM5 and TM7. The peak optical light contamination is localized at the bottom of the CCDs. This aspect could be used to select areas of the CCD where the thresholds could be decreased in order  to improve the low-energy performance. This could be done either by setting areas of the CCD to ``bad'' (using the bad pixel functionality) or by uploading a new threshold map to TM5 and TM7. Both approaches are possible but have implications in the operations of these cameras.
\end{itemize}
These approaches to mitigate the light leak effects will be analyzed, compared, and traded against each other during the second all-sky survey. Implications on added scientific return, overheads in operations, and mission planning will be considered for all mitigation options. We expect to have a plan of action for the start of the third all-sky survey, starting in December 2020.

\section{First light}

Following the extended commissioning phase discussed above, all seven TMs have been observing the sky simultaneously since October 15, 2019.
A field in the Large Magellanic Cloud (LMC) was chosen as a first light target, with the pointing centered on the supernova SN\,1987A.
Images were obtained in a series of exposures of all seven telescope modules with a combined integration time of about one day.

\begin{figure*}
\centering\includegraphics[width=18cm] {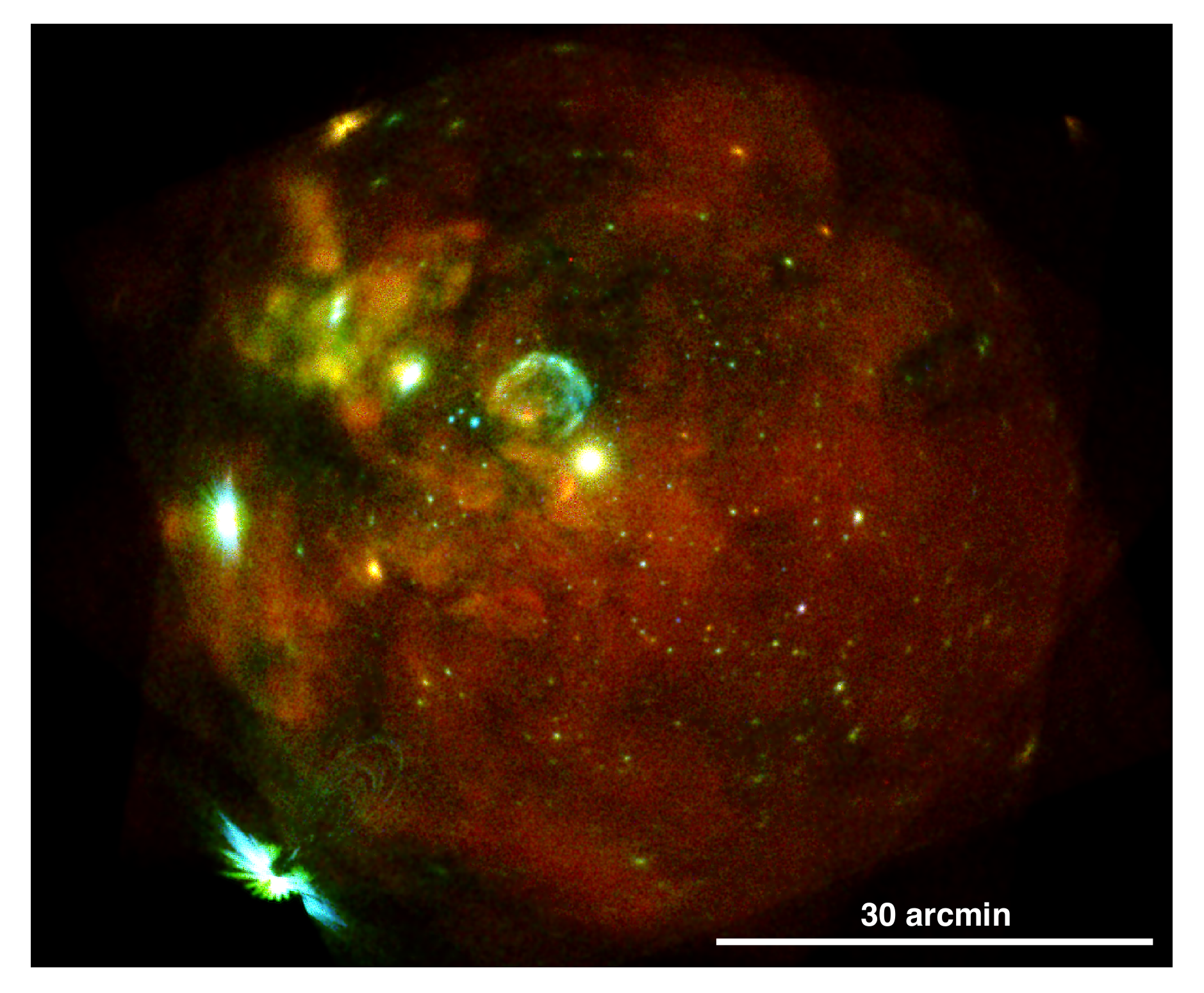}
\caption{False-color eROSITA X-ray image of the LMC region centered on the supernova SN\,1987A (the bright source which appears white-yellow,  southwest of the large shell structure 30 Doradus C). Red, green, and blue colors represent X-ray intensities in the 0.2--1.0, 1.0--2.0, and 2.0--4.5\,keV energy bands, respectively. The bright bluish feature in the southeast is caused by LMC\,X-1, the brightest X-ray source in the LMC. 
It was observed at large off-axis angle and covered by only a select few of the cameras.}
\label{fig:LMC}
\end{figure*}

In our neighboring galaxy (Fig.~\ref{fig:LMC}), eROSITA not only shows the distribution of diffuse X-ray emission from hot gas in this part of the LMC, but also some remarkable details, such as X-ray binaries and supernova remnants like SN\,1987A. SN\,1987A was first detected in the soft X-ray band with ROSAT in 1992 by \citet{1994A&A...281L..45B} and then rose in flux, first approximately linearly and then exponentially \citep{1996A&A...312L...9H,2006A&A...460..811H}, before reaching a maximum around the year 2014 \citep{2016ApJ...829...40F}. The eROSITA data of SN\,1987A now confirm that this source has started to decrease in brightness, indicating that the shock wave produced by the stellar explosion in 1987 now leaves the dense equatorial ring created by the progenitor star. In addition to a host of other hot objects in the LMC itself, eROSITA also reveals a number of foreground stars from our own Milky Way galaxy as well as distant AGNs, whose radiation pierces the diffuse emission of the hot gas in the LMC \citep[see e.g., ][]{2001A&A...365L.208H}.


\section{Outlook and Conclusions}

\begin{table*}
\caption{Summary of performance characteristics of the eROSITA telescope and its survey sensitivity. The background counts are based on the first all-sky survey data. For eRASS:1 the flux sensitivity in each band has been computed by taking all sources detected above a likelihood of 8 (soft band) or 10 (hard band), and measuring the flux below which the logarithmic number counts start declining. For the complete survey after four years (eRASS:1-8) the predictions are based on detailed simulations that include all instrumental effects and particle background intensity consistent with that measured at L2. For each field or region, we quote the total (un-vignetted) exposure in seconds. As discussed in the text, the corresponding effective (vignetted) exposures can be computed by dividing the total exposure by 1.88 and 3.31 for the soft and hard bands, respectively.
}             
\label{summary}      
\renewcommand{\arraystretch}{1.3}        
\begin{tabular}{|c|c|c|c|}     
\hline
    \multicolumn{2}{|c|}{}    & \multicolumn{2}{c|}{Energy Range} \\
    \cline{3-4}
    \multicolumn{2}{|c|}{} & Soft Band & Hard Band \\
    \hline
    \multicolumn{2}{|c|}{} & 0.2--2.3\,keV & 2.3--8\,keV \\
    \hline
    \multicolumn{2}{|c|}{FoV averaged effective area [cm$^2$]} & 1237 at 1keV & 139 at 5\,keV \\
    \hline
    \multicolumn{2}{|c|}{Total Background [10$^{-3}$ cts/s/arcmin$^2$]} & $\approx$ 3.7 & $\approx$ 2.1 \\
    \hline
  \multicolumn{4}{c}{{\it Point source sensitivity eRASS:1}} \\
  \hline
  Ecliptic Equatorial region & Total exposure = 200~s & $5 \times 10^{-14}$ erg/s/cm$^2$ & $7 \times 10^{-13}$ erg/s/cm$^2$ \\
  \hline
  Ecliptic Polar region & Total exposure = 4000~s & $7 \times 10^{-15}$ erg/s/cm$^2$ & $9 \times 10^{-14}$ erg/s/cm$^2$ \\
   \hline
  \multicolumn{4}{c}{{\it Point source sensitivity eRASS:1-8 (predicted)}} \\
  \hline
    Ecliptic Equatorial region & Total exposure = 1600~s & $1.1 \times 10^{-14}$ erg/s/cm$^2$ & $2.5 \times 10^{-13}$ erg/s/cm$^2$ \\
    \hline
  Ecliptic Polar region &  Total exposure = 30000~s & $2.5 \times 10^{-15}$ erg/s/cm$^2$ & $4 \times 10^{-14}$ erg/s/cm$^2$ \\
\hline
\end{tabular}
\end{table*}

eROSITA is a wide field, high-throughput X-ray telescope operating in the 0.2--8\,keV energy range. Its optical design is optimized for the detection of soft, diffuse X-ray emission from clusters of galaxies, which can be easily distinguished from point sources with the good PSF of the seven eROSITA telescopes. The design-driving scientific goal of eROSITA is the discovery of a sufficiently large number of clusters ($>$100\,000) to constrain cosmological models by studying the growth of the most massive structures in the Universe as a function of redshift. However, given the sensitivity of the telescope, and the all-sky nature of its observational program, the eROSITA X-ray data has an impact on the full breadth of astrophysical research, including for example compact objects, accretion physics, black-hole formation and growth, galaxy evolution, and stellar astronomy. The launch of eROSITA in summer 2019 from Baikonur onboard the Russian--German SRG mission represents a major milestone for the astrophysical community.

During its first year of operations in space, we have been able to validate most technical, operational, and scientific design characteristics of the eROSITA instruments on SRG. 
Table~\ref{summary} describes the main performance characteristics of eROSITA based on the data collected in this period, both during the performance verification phase and the all-sky survey. The performance closely  matches the expectations (see pre-launch estimates of \citep[][
Table 4.4.1 therein]{Merloni2012} in the soft-energy band, while it is slightly poorer in the hard band, mainly because of the increased level of particle background. 

In summary, we conclude that the data gathered during the calibration and performance verification phase and the first all-sky survey, while still being analyzed, instill confidence that the ambitious scientific goals of the mission can be met. A series of papers reporting on these early results are in preparation and will be presented elsewhere.

The eROSITA All-Sky Survey (eRASS), started on December 13, 2019, will continue until the end of 2023. As it takes SRG/eROSITA just six months to cover the whole sky, the survey will deepen progressively as the data are combined from its first pass (eRASS:1) to the eighth and final combined images of eRASS:1-8. The seven telescope modules of eROSITA which are scanning the sky will thus yield the deepest and sharpest 4$\pi$ survey ever made at soft (0.2-2.3~keV) energies, exceeding  the depth of the only previous all-sky imaging X-ray survey, ROSAT, by more than one order of magnitude. In the hard band (2.3-8~keV) they will provide the first ever true imaging survey of the sky. Considering the landscape of current and future X-ray missions, the eROSITA survey is likely to remain unparalleled for at least the next 15 years.

At the time of writing, as discussed above, the first all-sky survey pass has been completed. This, together with the subsequent seven passes, will constitute the main scientific goal and effort of the eROSITA science teams. Scientific projects and publications based on the eROSITA proprietary data are regulated by ``Working Groups''\footnote{The list of the German Consortium Working Groups is available at {\tt www.mpe.mpg.de/455860/working\_groups}}, with more than 150 scientists having already joined the various groups within the German Consortium only. In addition, a variety of individual and group external collaborations have been established, including a number of wide-area imaging and spectroscopic survey teams.

The German eROSITA Consortium plans regular public releases  of the half-sky eROSITA data over which it holds proprietary rights, consisting of X-ray source catalogs (including coordinates, counts, fluxes in various X-ray bands, morphological parameters) along with X-ray images, calibrated event files, and exposure maps. The current plan is to release all data from the first 6, 24, and 48 months of observations in 2022, 2024, and 2026, respectively. The data of the Russian Consortium will also be made public on a timescale that is still to be defined. 

Following the all-sky survey phase, SRG will continue operating in pointed and scanning mode, and the astronomical community will have the opportunity to use eROSITA and ART-XC via open calls for observing proposals.

\begin{acknowledgements}
eROSITA is the primary instrument aboard SRG, a joint Russian-German science mission supported by the
Russian Space Agency (Roskosmos), in the interests of the Russian Academy of Sciences 
represented by its Space Research Institute (IKI), and the Deutsches Zentrum für Luft- und Raumfahrt
(DLR). The SRG spacecraft was built by Lavochkin Association (NPOL) and its subcontractors, and is
operated by NPOL with support from IKI and the Max Planck Institute for Extraterrestrial Physics (MPE).
The development and construction of the eROSITA X-ray instrument was led by MPE, with contributions from
the Dr.~Karl Remeis Observatory Bamberg \& ECAP (FAU Erlangen-Nürnberg), the University of Hamburg 
Observatory, the Leibniz Institute for Astrophysics Potsdam (AIP), and the Institute for Astronomy and
Astrophysics of the University of Tübingen, with the support of DLR and the Max Planck Society. The Argelander Institute for Astronomy of the University of Bonn and the Ludwig Maximilians Universität 
Munich also participated in the science preparation for eROSITA. The eROSITA data shown here were
processed using the eSASS/NRTA software system developed by the German eROSITA consortium.
\newline
\newline
P.P. would like to express his deepest gratitude to all the colleagues in the team he has been working with closely for more than 10 years in order to let eROSITA become reality: Birgit Boller, Bernd Budau, Kurt Dietrich, Hans Eibl, Roland Gaida, Kati Hartmann, Johannes Hartwig, Franz Huber, Franz Oberauer, Christian Rohé, Thomas Rupprecht, Reiner Schreib, Fritz Schrey, Daniel Schuppe, and Franz Soller. You were a fantastic team! My thanks also go to my colleagues in the MPE technical departments and workshops, the administration for helping in personnel, financial and purchase order matters.

\end{acknowledgements}

\bibliographystyle{aa}
\bibliography{erositabib}

\end{document}